\documentclass[12pt]{iopart}
\expandafter\let\csname equation*\endcsname\relax
\expandafter\let\csname endequation*\endcsname\relax\usepackage{amsmath}
\usepackage{amssymb}
\usepackage{amsbsy}
\usepackage{graphicx}
\usepackage{bm}
\usepackage{color}
\usepackage[percent]{overpic}
\usepackage[colorlinks=true,citecolor=blue]{hyperref}
\hypersetup{colorlinks=true,citecolor=blue,linkcolor=red,urlcolor=blue}

\usepackage{fancyhdr}
\begin{document}

\title{Edge modes in zigzag and armchair ribbons of monolayer MoS$_2$}

\author{Habib Rostami}
\address{Istituto Italiano di Tecnologia, Graphene Labs, Via Morego 30, I-16163 Genova,~Italy}
\address{School of Physics, Institute for Research in Fundamental Sciences (IPM), Tehran 19395-5531, Iran}
\author{Reza Asgari}
\address{School of Physics, Institute for Research in Fundamental Sciences (IPM), Tehran 19395-5531, Iran}
\address{Condensed Matter National Laboratory, Institute for Research in Fundamental Sciences (IPM), Tehran 19395-5531, Iran}
\author{Francisco Guinea}
\address {IMDEA Nanociencia Calle de Faraday, 9, Cantoblanco, 28049, Madrid, Spain}
\address{School of Physics and Astronomy, University of Manchester, Oxford Road, Manchester M13 9PL, UK}

\ead{habib.rostami@iit.it}

\date{\today}
\begin{abstract}

We explore the electronic structure, orbital character and topological aspect of a monolayer MoS$_2$ nanoribbon using tight-binding (TB) and low-energy (${\bm k}\cdot{\bm p} $) models. We obtain a mid-gap edge mode in the zigzag ribbon of monolayer MoS$_2$, which can be traced back to the topological properties of the bulk band structure. Monolayer 
MoS$_2$ can be considered as a valley Hall insulator. The boundary conditions at armchair edges mix the valleys on the edges, and a gap is induced in the edge modes. The spin-orbit coupling in the valence band reduces the hybridization of the bulk states.

\end{abstract}
\maketitle

%%%%%%
\section{Introduction}\label{sec:intro}
%%%%%%
The electronic properties of two-dimensional (2D) crystalline materials can be efficiently tuned by their edge structures and thus the 2D nanoribbons can show various functional features, including metallic, semi-metallic, semiconducting, and magnetic~\cite{lin12, kang12, pan12, wang10}.
The most straightforward method to study finite size effects is a lattice model in real space, and a practical example is given by \textit{ab-initio} simulations. Besides, a tight-binding method could also be a powerful lattice-based model in order to survey the effect of different boundary terminations in ribbon cases. Those models have been extensively used for graphene nanoribbons~\cite{Cai14} where partly flat band edge modes in a ribbon geometry exist \cite{Ryu02, Eduardo_2008,Yao09}.

The MoS$_2$ nanoribbons can be directly obtained by cutting the MoS$_2$ monolayer and according to the directions of termination, there exist two well-known kinds of nanoribbons namely; armchair and zigzag.
Although some other kinds of edge termination like S-dimmer, S-half, S$_2$-strip and antisymmetric edge structures might be possible to make and to exist \cite{ES12}, the most symmetric edge structure are the armchair and zigzag edge terminations.
From \textit{ab-initio} calculations, a zigzag ribbon of monolayer MoS$_2$ reveals metallic edge states closing the gap \cite{BB01,BN03,LC08,CZ14,GM15} while, an armchair ribbon of monolayer MoS$_2$ contains gapped edge modes \cite{LC08,AC11,DS12}. Intriguingly, the metallic edge modes of monolayer MoS$_2$ have been also observed experimentally for a MoS$_2$ triangular nanocrystal on Au (111) \cite{HB00} in which the existence of edge modes on the zigzag termination of the nanocrystal of MoS$_2$ is found in atomic-resolved STM image. Moreover, very recently the metallic phase of the edge modes in monolayer MoS$_2$ on graphite has been illustrated through a finite conductivity in the gap region, and also by STM spectroscopy \cite{CC14}. Recently, the contribution of these edge modes in the non-linear optical properties (e.g. second harmonic generation) of MoS$_2$ is 
observed experimentally \cite{edge_shg} which manifests a novel importance of the edge modes in this electronic 2D system.
These experimental evidences imply that these metallic edge modes are robust versus disorder therefore it is worth to explore them in a theoretical manner as well.
These features make them promising for the application in electronics and optoelectronics \cite{ZH13,HT13}. 
We would like to notice that these edge modes are formed due to the specific termination of the edges, however, edge modes can be also artificially created by using external gating \cite{AY12} mimicking a hard-wall boundary condition on the effective edge. Recently, the boundary condition of the transition metal dichalcogenides (TMDs) ribbon has been studied \cite{PKB15} by using a ${\bm k} \cdot {\bm p}$ model \cite{KF15} and following M-matrix approach \cite{MF04,AB08}.

The topological nature of the metallic edge states in zigzag ribbons and the origin of the gapped edge modes in armchair one have not been discussed in literature. The focus of this article is the understanding of these issues. We will use a tight-binding model \cite{CG13,CS13,RO14,RA15} which provides quite accurate dispersion relations in $\bf k$-space. The reflection symmetry around the central plane of the layer allows us to split the hamiltonian into an even and odd sector. The symmetry ($\sigma_h$) maps $z$ to $-z$. The low-energy band structure of the system belongs to the even subspace \cite{CG13}. In this even subspace, the symmetry of the system and the model is very similar to the case of gapped graphene and one can expect some similarity between these two systems.
The topological aspect of the low-energy Hamiltonian of monolayer MoS$_2$ have been discussed before \cite{RMA13,RA14,RG15} where a possible non-trivial phase according to the non-zero Chern number for each spin and valley flavor has been addressed. The low-energy Hamiltonian of this system is not exactly a massive Dirac model, as it includes a momentum dependent mass ($\Delta+\beta q^2$) similar to the modified Dirac model of a topological insulator thin film \cite{Shen12} and to the HgTe-based quantum spin Hall system (BHZ-model) \cite{BHZ06}. Some of the topological aspects of a modified-Dirac system have been explored after discretizing the low-energy model on a square lattice using a finite difference (FD) method \cite{IK10}.

In this paper, we explore the edge state dispersion of the zigzag and armchair ribbons of monolayer MoS$_2$ as the main representative of TMDs using a tight-binding model. The numerical calculations are supported by using a low-energy continuum model. We obtain metallic and non-metallic edge modes for the zigzag and armchair ribbons, respectively.
We address as well the topological aspects of the low-energy model of monolayer MoS$_2$ through numerical calculations in a finite size system and analytical arguments. A finite element method is employed to discretize the two-band continuum model on a ribbon geometry. We describe that the weak topological nature of monolayer MoS$_2$ originates from the particular orbital character of each energy band and it is not related to the spin-orbit coupling in contrast with graphene. We calculate the Berry curvature, valley Chern number, and Z$_2$ invariant, in order to discuss the topological nature of the continuum model. The topological relevance of  the trigonal warping is also taken into account.
Two effective one-dimensional (1D) Hamiltonians are introduced to describe the most relevant edge modes in both the armchair and zigzag ribbons. We find a 1D massless (massive) Dirac cone for the edge modes of the zigzag (armchair) ribbon. The gapped spectrum of the edge modes in the armchair ribbon arises from the combination of the effects from $\beta q^2$ mass-term in the continuum ${\bm k}\cdot {\bm p}$ model and a hybridization between 1D modes on two edges of the ribbon. This hybridization can be also understood in terms of a mixing of 1D-valleys (Dirac cones) on the edges.

This paper is organized as follows. Sec. \ref{sec:theory-model} describes the models and approximations used. In Sec. \ref{sec:TB} and Sec. \ref{sec:KP} our main results concerning the edge mode dispersions of MoS$_2$ nanoribbons are shown and discussed. In Sec. \ref{sec:topol}, we study the topological aspect of the system by the calculating Berry curvature, the Chern number and the Z$_2$ invariant. Finally, we conclude our main results regarding  the finite size effects of the systems in Sec. \ref{sec:conclusion}.

%%%%%%
\section{Theory and model}\label{sec:theory-model}
\begin{figure}[h]
 \centering
\includegraphics[width=0.65\linewidth]{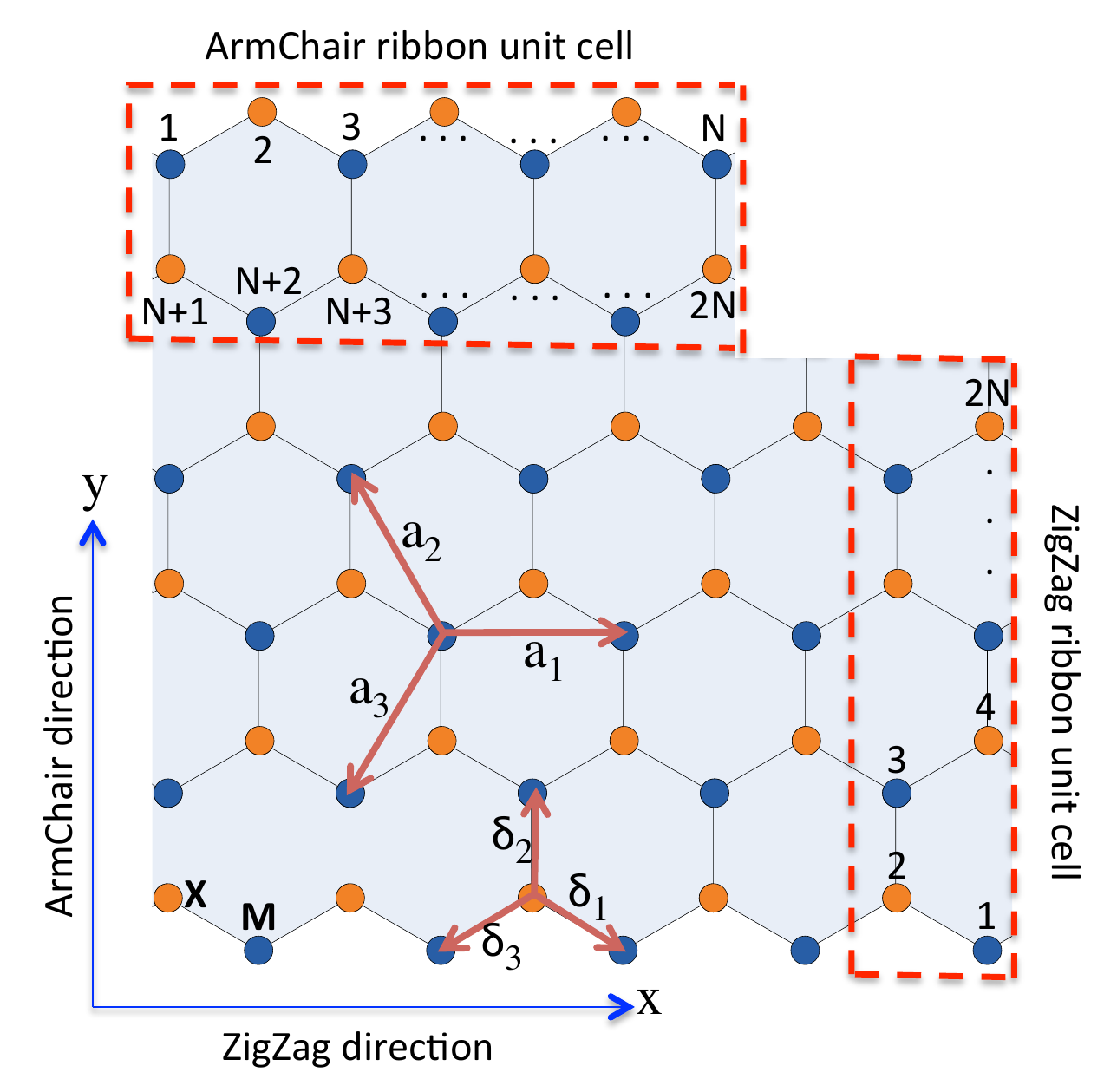}
\caption{(Color online)  Lattice structure of monolayer $MX_2$ in which the zigzag and armchair directions and corresponding unit cells are illustrated. The direction of the nearest (${\bm \delta}_i$) and the next nearest  (${\bm a}_i$) neighbour vectors are depicted on the lattice.}
\label{fig:scheam}
\end{figure}

\begin{figure}[h]
\centering
\includegraphics[width=0.6\linewidth]{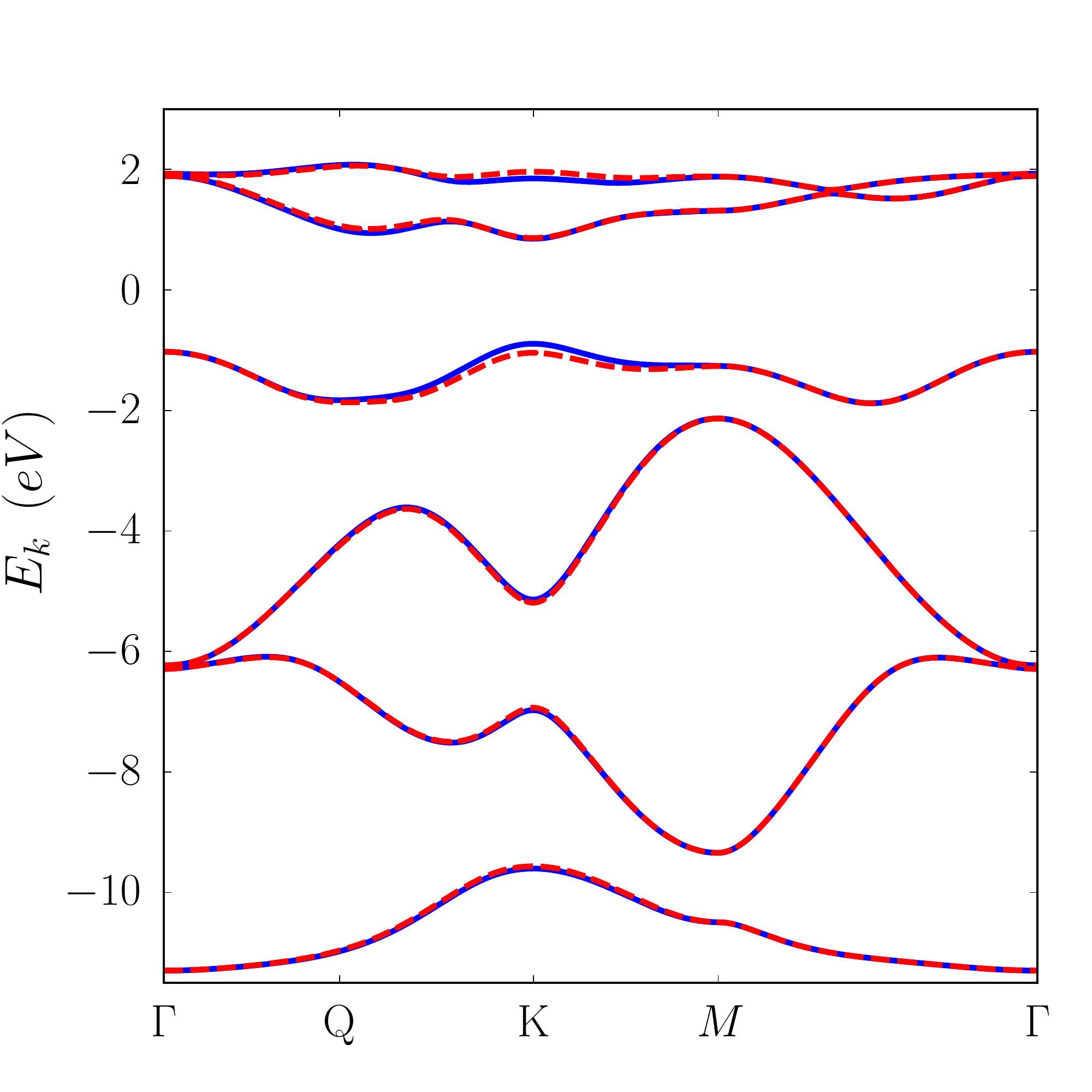}
\caption{(Color online) Band dispersion along ${\rm \Gamma-Q-K-M-\Gamma}$ direction for both spin components which are indicated by solid and dashed curves for the up and down component of spin, respectively.}
\label{fig:TB_dispersion}
\end{figure}

\begin{figure}
\centering
\begin{overpic}[width=0.7\linewidth]{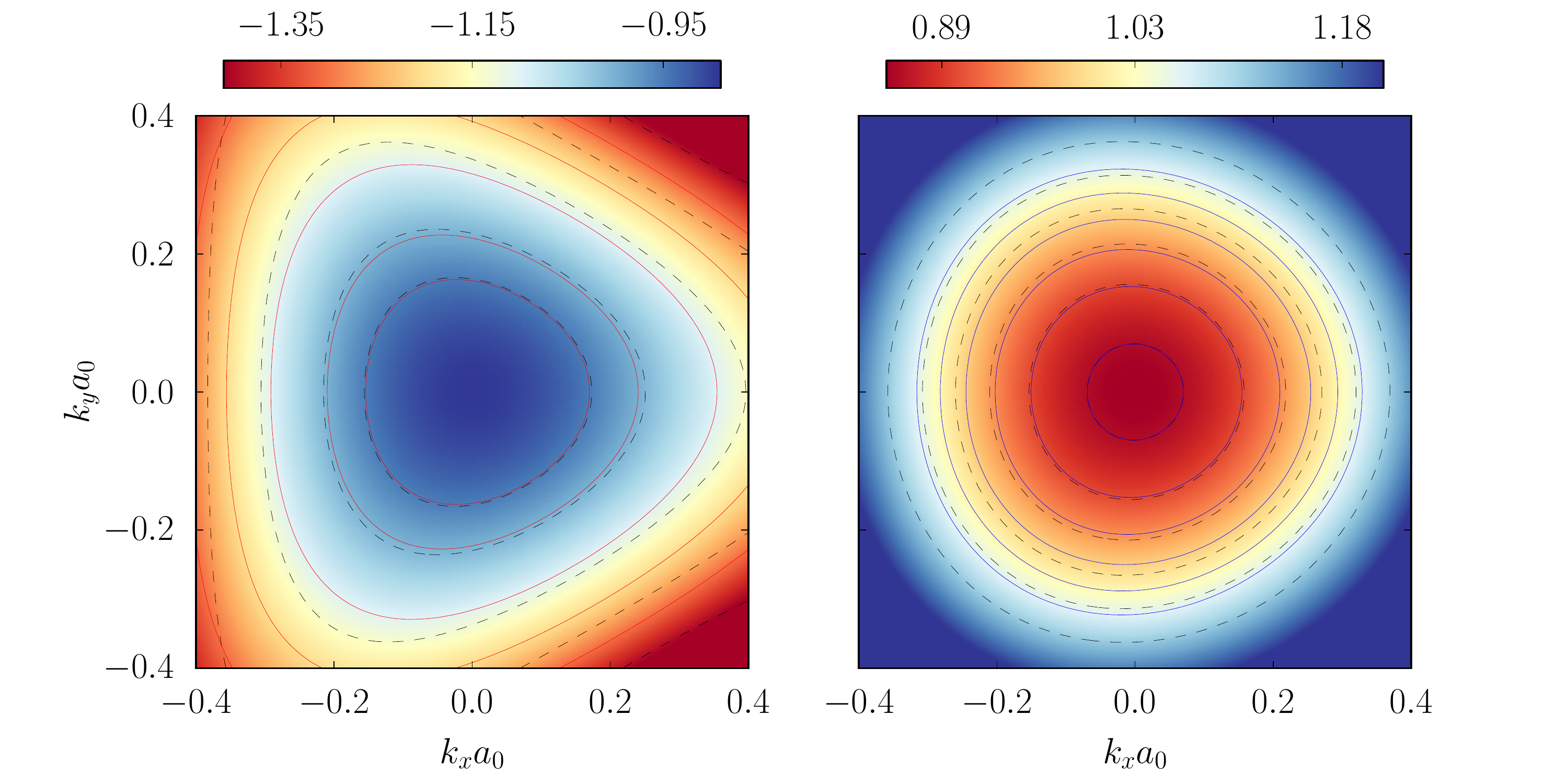}\put(0.0,43){(a)}\end{overpic}
\begin{overpic}[width=0.7\linewidth]{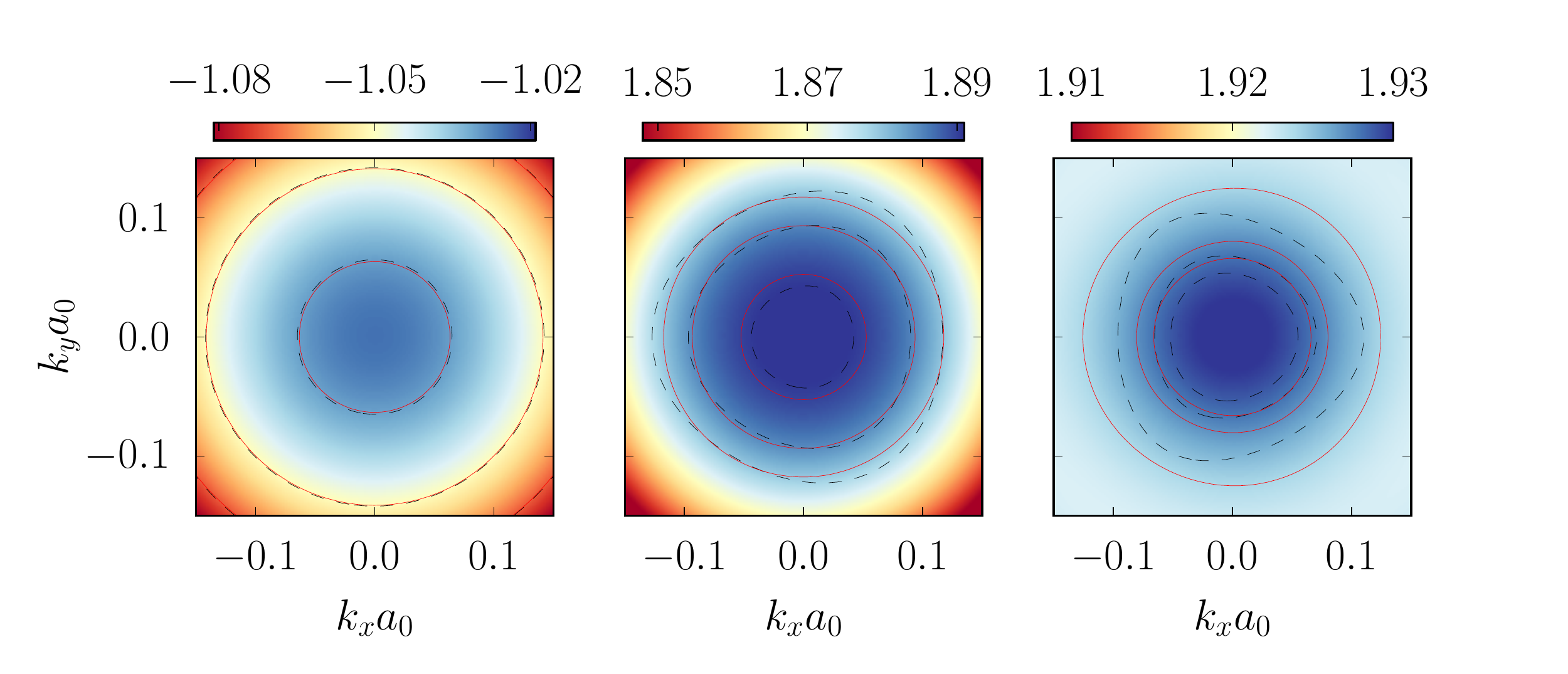}\put(0.0,35){(b)}\end{overpic}
\caption{(Color online) Comparison between the low-energy and six-band models around K and $\rm \Gamma$ points in (a) and (b) panels, respectively. 
Solid (dashed) contour line of energy dispersion indicates the results of two (six) band Hamiltonian. 
One can simply recognize the conduction (with positive energy) and valence band (with negative energy) through the negative and positive numbers indicating the energy levels on the color bars.}
\label{fig:kp_dispersion}
\end{figure}

%%%%%%
In order to study the finite size effect in monolayer TMDs, $MX_2$, where $M$ refers to a metal and $X$ indicates the chalcogen atom compound, we use a Slater-Koster tight-binding model which captures the main energy bands of this system in the whole Brillouin Zone (BZ).  This model has been already used to investigate the valley Zeeman effect and the effect of strain on a MoS$_2$ monolayer \cite{RA15, RG15}.  This tight-binding (TB) model for the single layer TMDs, uses an orbital basis which includes $\left(d_{3z^2-r^2},~d_{x^2-y^2},~d_{xy},p_{x}^{\rm sym},~p_{y}^{\rm sym},~p_{z}^{\rm antisym}\right)$, which are even w.r.t. the horizontal reflection symmetry ({\it i.e.} $\sigma_h:z\rightarrow -z$) of monolayer $MX_2$. Notice that the $d$ and $p$ orbitals in this model belong to the metal ($M$)
and the chalcogen ($X$) atoms, respectively. To clarify the {\it symmetric} and {\it antisymmetric} hybridisation of chalcogen $p$-orbitals localized on top and bottom layers, we use $2 p^{\rm sym}_{x,y}=p^{\rm top}_{x,y}+p^{\rm bottom}_{x,y}$ and $2 p^{\rm antisym}_z=p^{\rm top}_z-p^{\rm bottom}_z$. Spin-orbit coupling is treated in an atomic approximation as $\propto L_z S_z$ which describes well the spin-splitting of the valence band in this system.  The TB Hamiltonian in position space reads
%%%%%%
\begin{equation}
\label{Eq:H-Real}
{\cal H}^{\rm TB}
=
\sum_{i,\mu\nu}
\epsilon_{\mu,\nu} c^\dagger_{i,\mu}c_{i,\nu}
+
\sum_{ ij,\mu\nu}
{[t_{ij,\mu\nu} c^\dagger_{i,\mu}c_{j,\nu}+{\rm H.c.}]}
\end{equation}
%%%%%%
where $c^\dagger_{i,\mu}$ creates an electron in the $i^{\rm th}$ unit cell in the atomic orbital labelled by $\mu$. All of the details corresponding to the hopping and on-site matrices can be found in  Ref.~[\cite{RG15}].
In Fourier space, we obtain a six-band Hamiltonian for each spin degree of freedom as function of wave vector (${\bm k}$) and spin ($s$) which is given by
%%%%%%
\begin{equation}\label{eq:HTBk}
{\cal H}^{\rm TB}_s({\bm k})=\epsilon_s+
\sum_{i=1}^{3}\begin{bmatrix}2t_i^{MM}\cos \left({\bm k}\cdot{\bm a}_i\right)&&
 {t^{MX}_i e^{-i {\bm k}\cdot{\bm \delta}_i}}
 \\
 {t^{XM}_i e^{i {\bm k}\cdot{\bm \delta}_i}}&&2t_i^{XX}\cos \left({\bm k}\cdot{\bm a}_i\right)\end{bmatrix}
\end{equation}
%%%%%%
where $\epsilon_s={\rm diag}\left[\epsilon^{M}_s,\epsilon^{X}_s\right]$
and the nearest (${\bm \delta}_i$) and the next nearest (${\bm a}_i$) neighbour vectors are shown in Fig. \ref{fig:scheam}.
 %%%%%%
 %%%%%%
The hopping terms, $t_{ij,\mu\nu}$, are evaluated within a Slater-Koster scheme~\cite{CS13,RG15}. After diagonalizing the Hamiltonian, the band dispersion along the ${\rm\Gamma -Q -K -M -\Gamma}$ direction is calculated and the results are given in Fig. \ref{fig:TB_dispersion} which shows the main energy bands of the system which belong to the {\it even} subspace of the orbital basis with respect to the horizontal reflection symmetry.
It should be mentioned that the $Q$ point, which is located midway between $\rm\Gamma$ and K points, is not a high symmetry point, albeit it is where one of the minima in the conduction band. The anisotropic effects, {\it i.e.} trigonal warping in the low-energy bands, are illustrated in Fig. \ref{fig:kp_dispersion} for both conduction and valence bands around the K and $\rm\Gamma$ points as discussed in detail below.
%%%%%%
%%%%%%
The conduction band at the K-point is mostly made from $d_{z^2}$ orbital with some mixing from $p_x$ and $p_y$. The valence band contains solely $d_{xy}$ and $d_{x^2-y^2}$ orbitals. At the $\rm\Gamma$ point, the valence band is mainly built from the $d_{z^2}$ orbitals with a minor contribution from $p_z$ orbitals. The two nearly degenerate conduction bands in $\rm\Gamma$ point contain a combination of four other orbitals with equal weight.

Having performed a canonical perturbation on the Hamiltonian given in (\ref{eq:HTBk}), we can deduce a low-energy two-band model around K-points from the six-band tight-binding model for each spin component, which can be written as ${\cal H}_{\tau s}={\cal H}^i_{\tau s}+{\cal H}^w_{\tau s}$ where \cite{RG15}
%%%%%%
\begin{gather}\label{eq:HkpK}
{\cal H}^i_{\tau s}({\bm q}) = \frac{\Delta_0+\lambda_0\tau s}{2}+\frac{\Delta+\lambda\tau s}{2}\sigma_z+t_0 a_0 {\bm q}\cdot{\sigma}_\tau
\nonumber\\
\hspace{1.7cm}+\frac{\hbar^2|{\bm q}|^2}{4m_0}(\alpha+\beta\sigma_z)
+a_0^2|{\bm q}|^2(\lambda_0'+\lambda'\sigma_z)\tau s ~, 
\nonumber \\
{\cal H}^w_{\tau s}({\bm q}) = t_1 a_0^2{\bm q}\cdot{\sigma}^{\ast}_{\tau}\sigma_x {\bm q}\cdot{\sigma}^{\ast}_{\tau}
+t_2a_0^3\tau(q_x^3-3q_xq_y^2)(\alpha'+\beta'\sigma_z)~.
\end{gather}
%%%%%%
where $s=\pm$ and $\tau=\pm$ stand for the spin and valley degree of freedom, respectively. Notice that ${\sigma}_{\tau}=(\tau\sigma_x,\sigma_y)$ with $\sigma_{i=x,y,z}$ are Pauli matrices, ${\bm q}=(q_x, q_y)$ is the wave vector in two dimensions, $m_0$ is the free electron mass and $a_0=a/\sqrt{3}$ where $a\approx 3.16$~\AA~is the lattice constant. The numerical values of two-band model parameters are $\Delta_0=-0.11$~eV, $\Delta=1.82$~eV, $\lambda_0=69$~meV, $\lambda=-80$~meV, $\lambda'_0=-17$~meV, $\lambda'=-2$~meV,  $t_0=2.34$~eV, $\alpha=-0.01$, $\beta=-1.54$, $t_1=-0.14$~eV, $t_2=1$~eV, and $\alpha'=0.44$, $\beta'=-0.53$. The total spin-orbit coupling is $\lambda_{\pm}=(\lambda_0\pm\lambda)/2+a^2_0|{\bm q}|^2 (\lambda'_0\pm\lambda')$ where $+ (-)$ stands for the conduction (valence) band which also has a quadratic momentum dependence.\par
${\cal H}^w_{\tau s}$ stands for the trigonal warping (TW) in both conduction and valence bands. The cubic terms in $q$ are similar to the cubic terms obtained previously in a three-band tight-binding model \cite{LX13}. We check the validity range of the low-energy Hamiltonian by comparing the energy dispersion of the low-energy model with that of six-band model around the K-point in Fig. \ref{fig:kp_dispersion}. The two sets of energy bands  coincide satisfactorily around the K-point. Using this Hamiltonian, we calculate the effective masses of the conduction and valence bands as $m_e=0.513m_0$ and $m_h=-0.503m_0$. The values for the effective masses are consistent with the negative sign of $\alpha$ parameter, which stands for the mass asymmetry between the conduction and valence bands.\par
To find the role of different parameters in the trigonal warping direction and strength, we calculate the low-energy band dispersion around the K points
%%%%%%
$E^{\pm}_{\tau s}({\bm q})=A_{\tau s}({\bm q})\pm\sqrt{B_{\tau s}({\bm q})^2+C({\bm q})}$ where
\begin{gather}
A_{\tau s}({\bm q})= \frac{\Delta_0+\lambda_{-}\tau s}{2}+2b\alpha|a_0{\bm q}|^2+t_2\alpha'|a_0{\bm q}|^3\cos(3\phi)~, \nonumber\\
B_{\tau s}({\bm q}) \frac{\Delta-\lambda_{-}\tau s}{2}+2b\beta|a_0{\bm q}|^2+t_2\beta'|a_0{\bm q}|^3\cos(3\phi)~, \nonumber\\
C({\bm q})=t_0^2|a_0{\bm q}|^2+t_1^2|a_0{\bm q}|^4+2t_0t_1|a_0{\bm q}|^3\cos(3\phi)~.
\end{gather}
%%%%%%
in which $b=\hbar^2/4m_0a_0^2\approx 0.572$eV. Notice that $\lambda_{+}$ and the momentum dependence of the spin-orbit coupling are neglected in these equations. The trigonal warping arises from the three parameters ($\alpha',\beta',t_1$) and all of these terms can be combined to obtain a term $z_{\pm}\cos(3\phi)$ in the low-energy dispersion where $z_{\pm}=t_2(\alpha'\pm\beta')\pm2 t_0 t_1/(\Delta-\lambda_{-}\tau s)$. Here, $z_+(z_-)$ stand for the conduction (valence) band. For small $q$'s, the direction of warping in both bands is different if $z_{+} z_{-}>0$. They are in the same direction when $z_ {+} z_ {-} <0$. If one of these parameters is zero, the corresponding band is isotropic. In the case that $\alpha'=0$, the warping in both bands are in the same direction and with the same warping strength. The parameters $\alpha$ and $\alpha'$ are the sources of asymmetry in the effective masses and they define the trigonal warping directions in the conduction and the valence band.  We obtain that $z_{+} z_{-}<0$, which means same warping direction in two bands which can be seen clearly in Fig. \ref{fig:kp_dispersion} around the K-point.
%%%%%%
%%%%%%
We have also extracted a low-energy model Hamiltonian around the $\rm\Gamma$ point. Using L\"{o}wdin partitioning, we obtain a low-energy three-band Hamiltonian around the $\rm\Gamma$ point, including two degenerate bands in the conduction plus one band in the valence, and it can be written as follows
%%%%%%
\begin{equation}\label{eq:HkpG}
{\cal H}^{\Gamma}({\bm q})=\begin{bmatrix}\Delta^{\Gamma,1}_c+\gamma_3 |{\bm q}|^2 &&\gamma_1 q^\ast&&\gamma_2 q\\ \gamma_1 q&&\Delta^{\Gamma,2}_c+\gamma_3 |{\bm q}|^2&&\gamma_2 q^\ast\\ \gamma_2 q^\ast&&\gamma_2 q&&\Delta^\Gamma_v+\gamma_4 |{\bm q}|^2\end{bmatrix}\nonumber\\
\end{equation}
%%%%%%
where $q=q_x+i q_y$, $\Delta^{\Gamma,1}_c=(1.91+0.02s)$~eV, $\Delta^{\Gamma,2}_c=(1.91-0.02s)$~eV, $\Delta^\Gamma_v=-1.02$~eV, $\gamma_1 a_0=-0.0005s$~eV, $\gamma_2 a_0=1.518$~eV, $\gamma_3 a_0^2=-1.437$~eV, $\gamma_3 a_0^2=-1.45$~eV, and $\gamma_4 a_0^2=0.32$~eV. It should be mentioned that, at the $\rm\Gamma$ point, we have not considered higher order terms in $q$ to capture the trigonal warping in two conduction bands. The average over spin of the $\gamma_1$ gives zero and this coefficient can be neglected. Using this Hamiltonian, we find that the effective masses are $m^{(1)}_c=-1.77m_0$, $m^{(2)}_c=-1.73m_0$ and $m_v=-0.91m_0$ for two conduction and one valence bands, respectively. A comparison between the three and six band model around the $\rm\Gamma$ point is shown in Fig. \ref{fig:kp_dispersion} in the bottom panels.

%%%%%%
\section{Zigzag and armchair ribbons: Tight-binding model}\label{sec:TB}
%%%%%%

\begin{figure}
\centering
\includegraphics[width=0.49\linewidth]{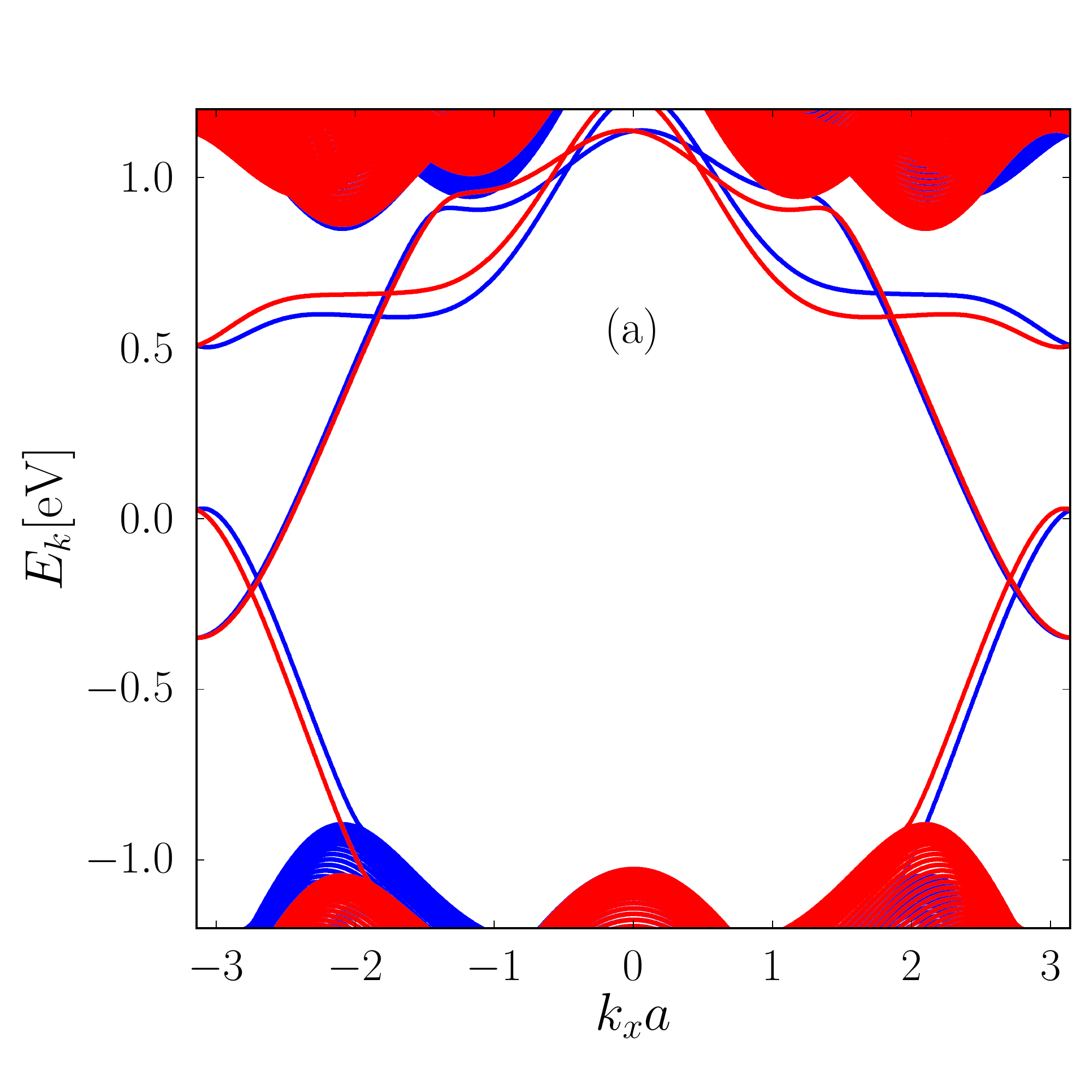}
\includegraphics[width=0.50\linewidth]{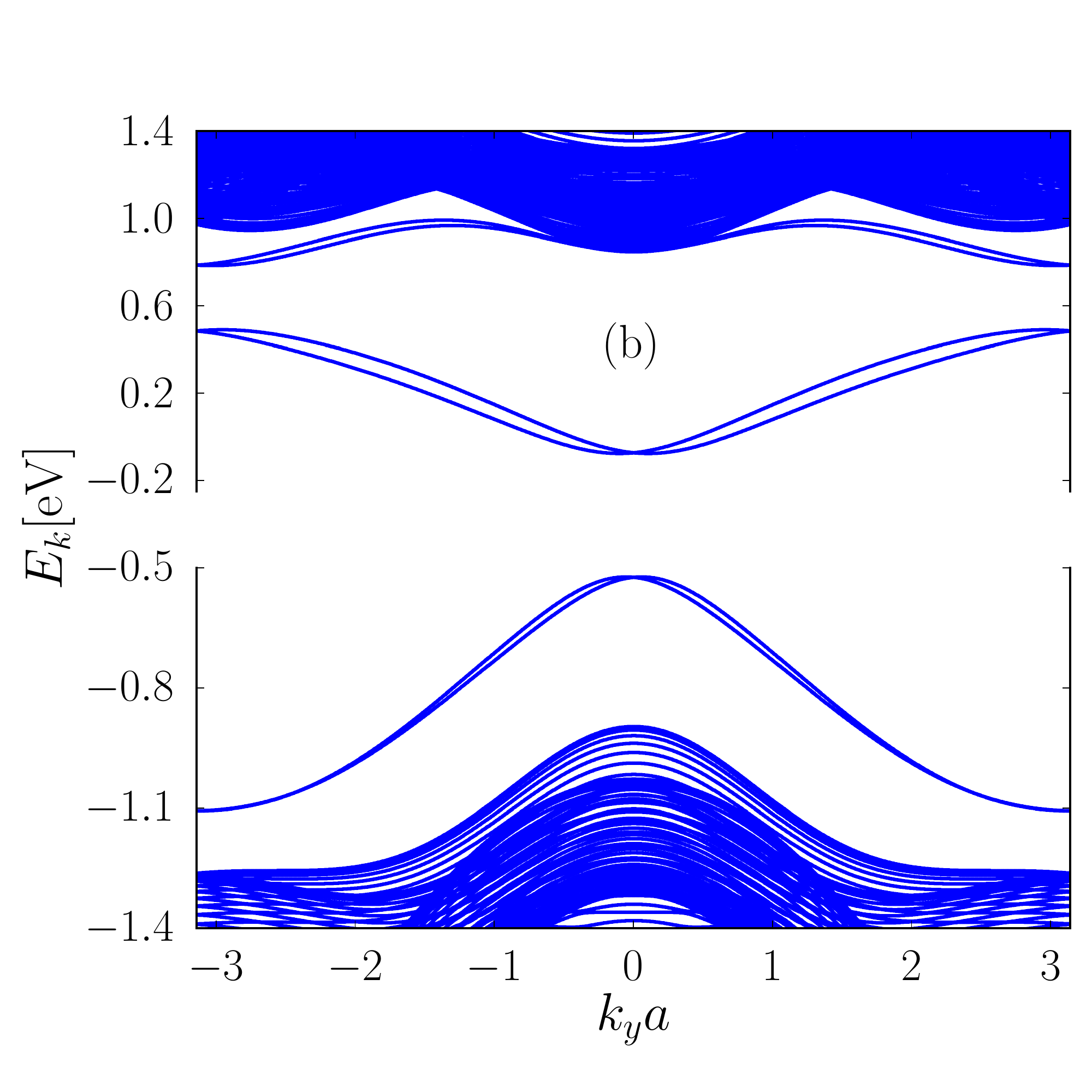}
\caption{(Color online) Energy dispersion of MoS$_2$ ribbons. (a) Zigzag ribbon N=100. (b) Armchair ribbon N=101. Red and blue colors indicate spin components. Owing to the spin degeneracy in the armchair ribbon just spin up component is plotted. The definition of $N$ for both zigzag and armchair ribbons is depicted on the Fig. \ref{fig:scheam}.}
\label{fig:TB-ribbon-dispersion}
\end{figure}

\begin{figure}[h!]
\centering
\includegraphics[width=0.55\linewidth]{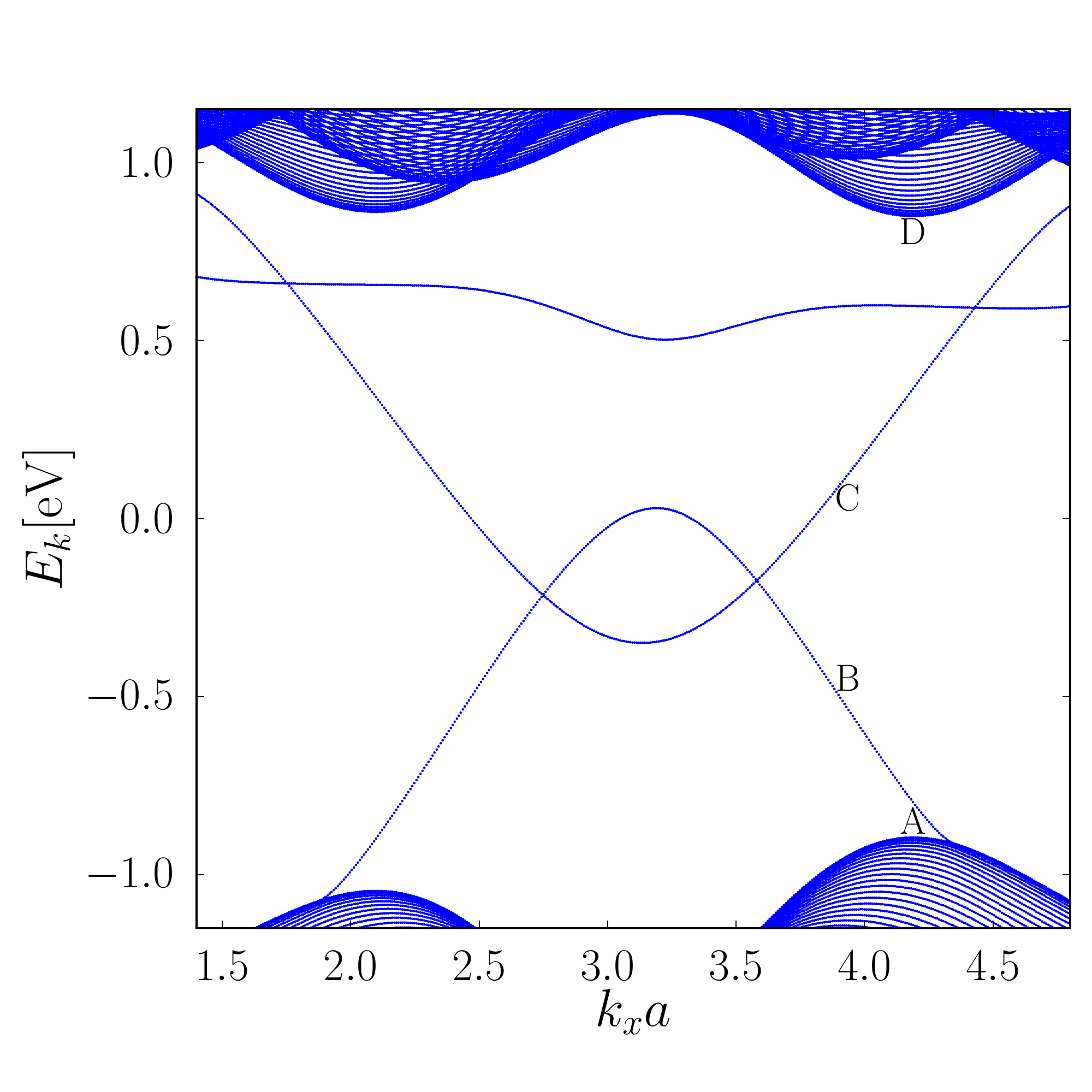}
\\
\includegraphics[width=0.35\linewidth]{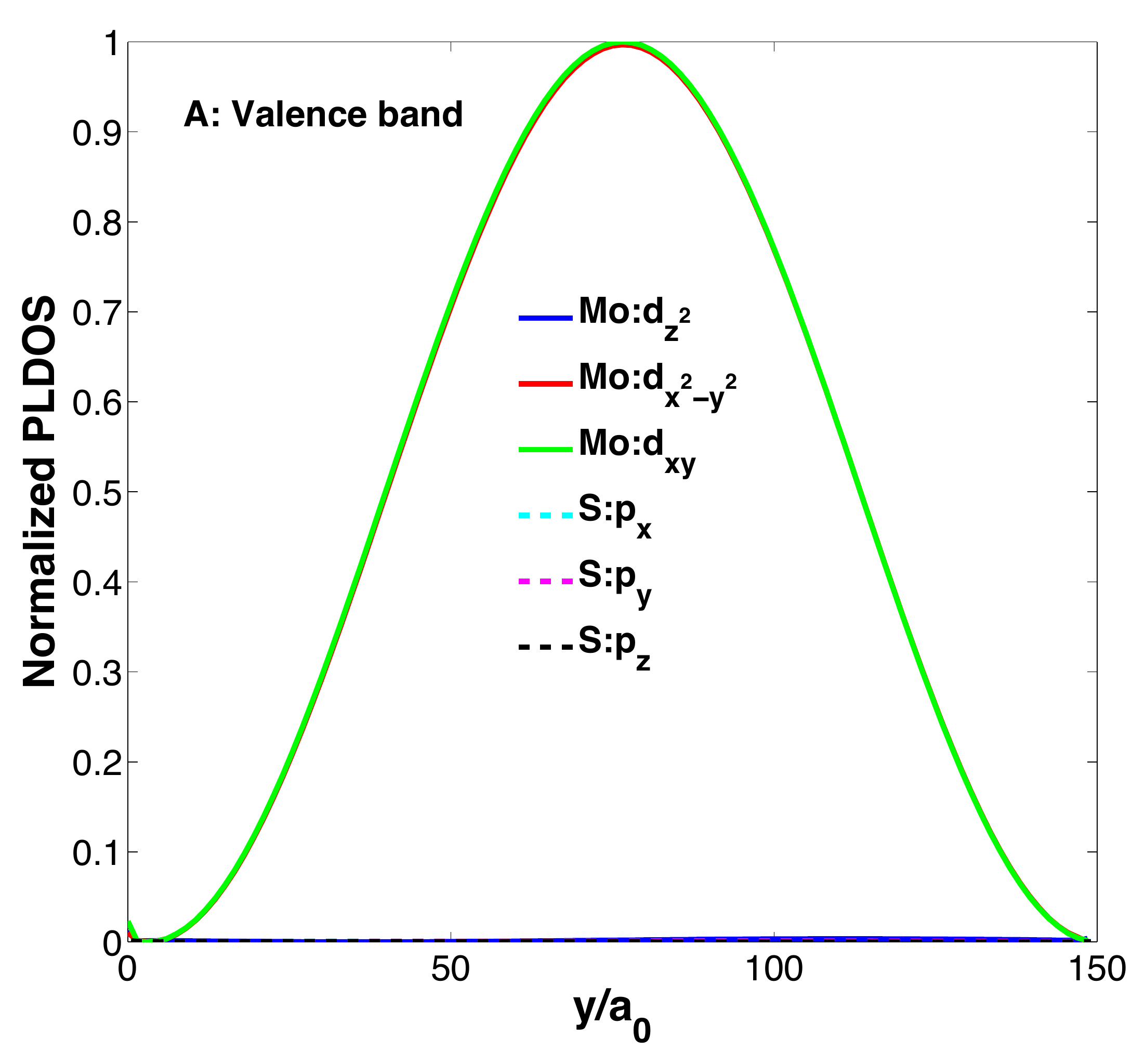}
\includegraphics[width=0.34\linewidth]{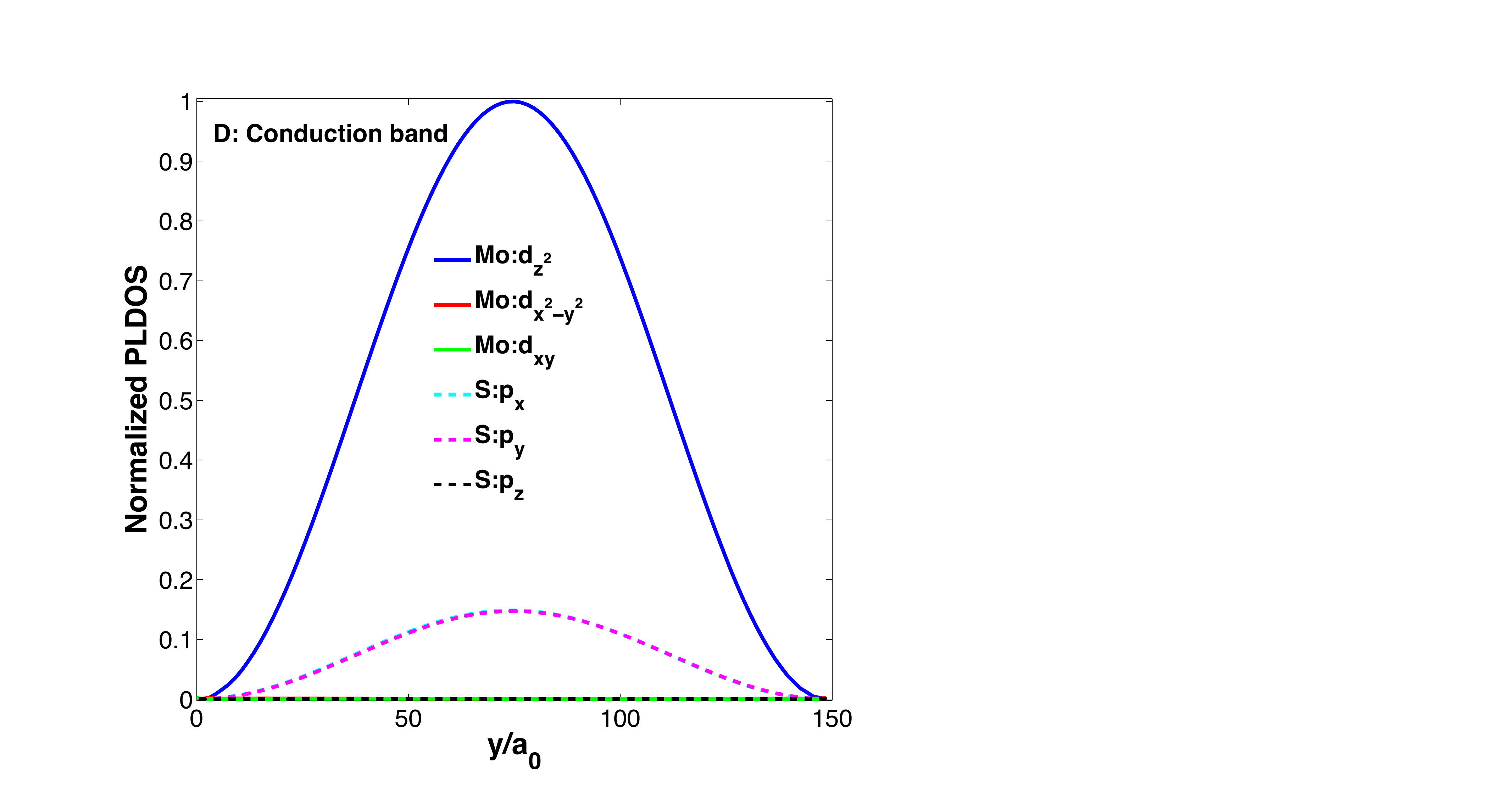}
\includegraphics[width=0.35\linewidth]{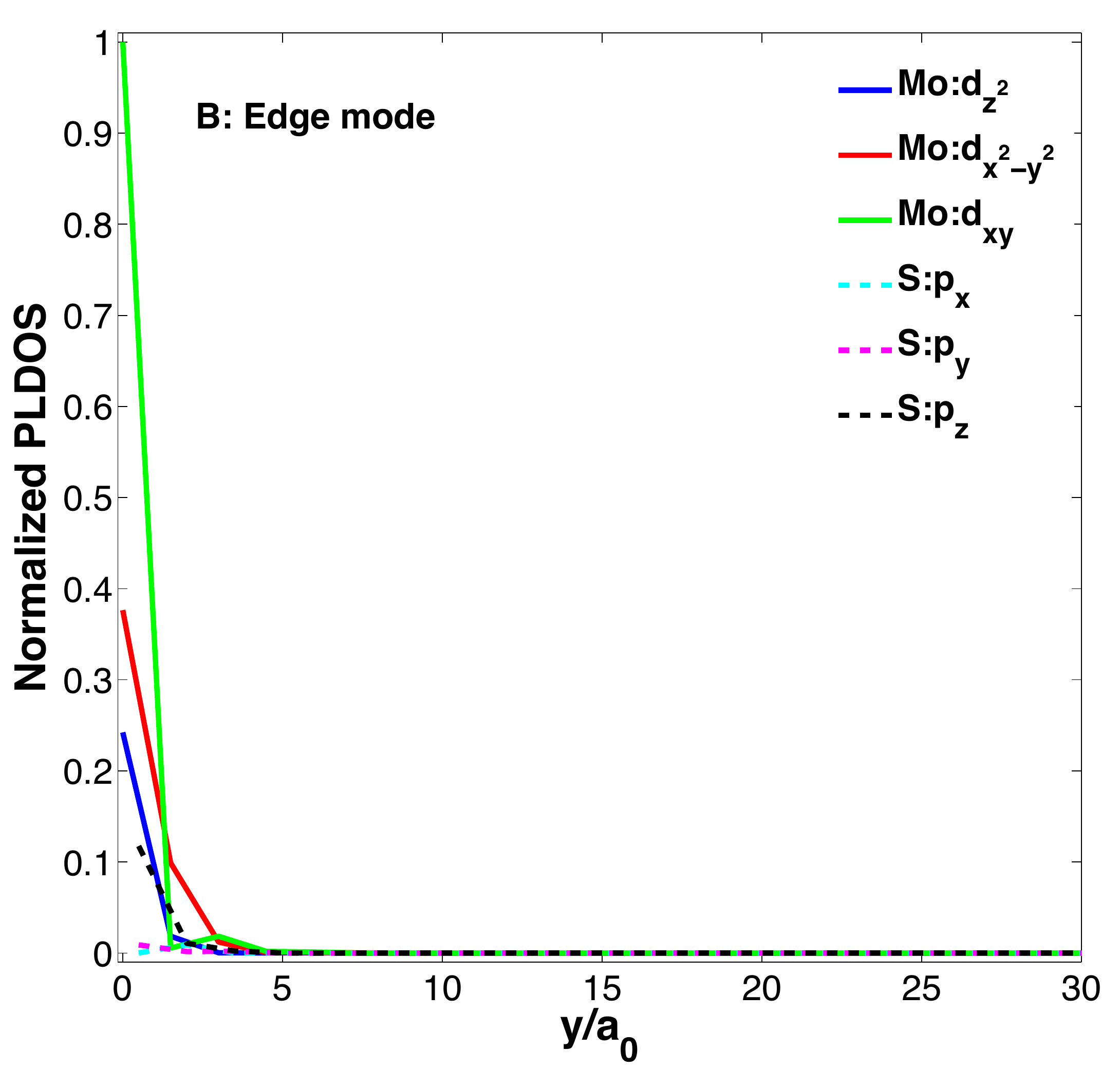}
\includegraphics[width=0.34\linewidth]{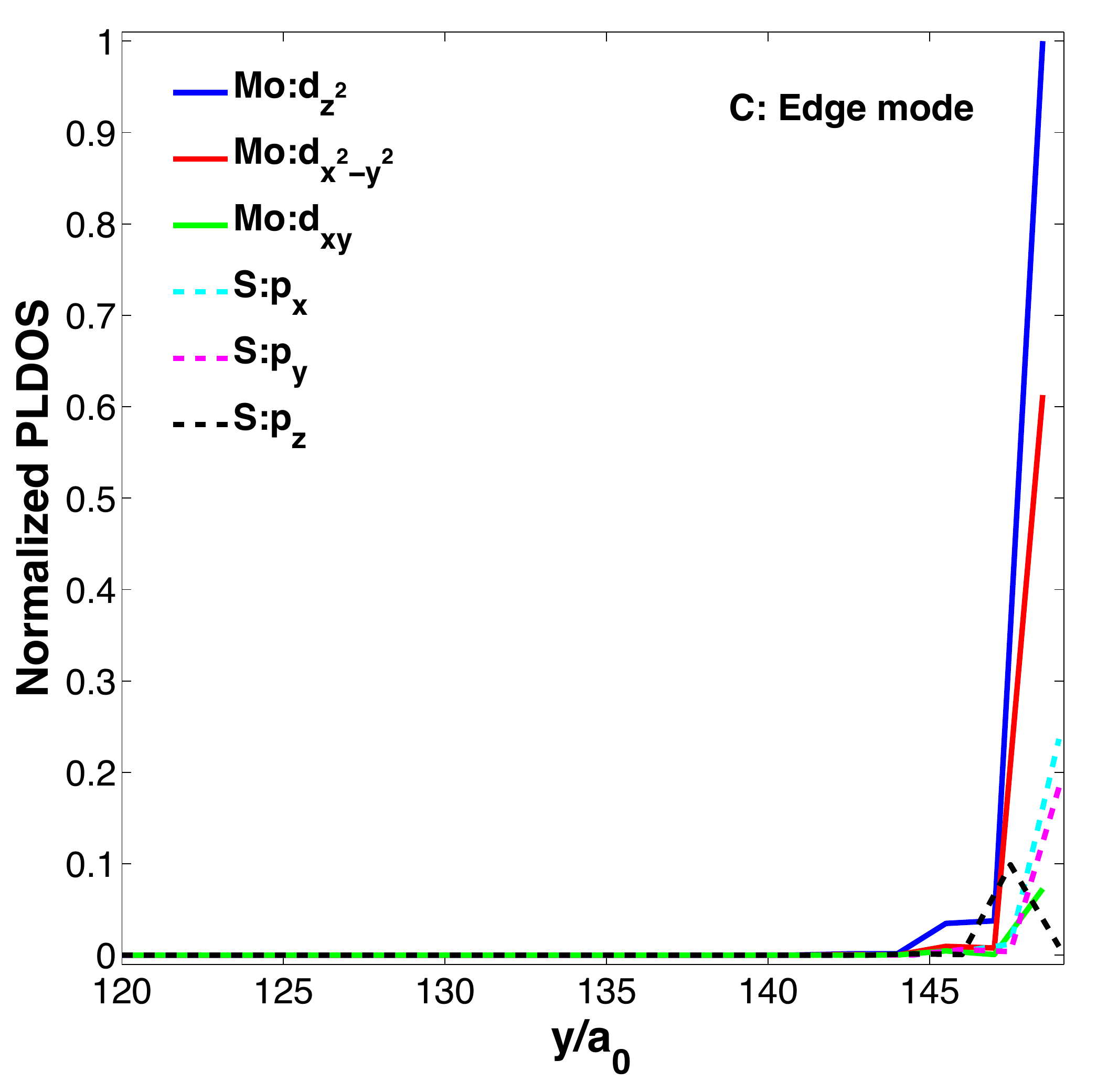}
\caption{(Color online) Projected local density of state for some spin-up states (shown in the first panel) of the zigzag ribbon with $N=100$ in the presence of the spin-orbit coupling.}
\label{fig:zig}
\end{figure}

Tight-binding model is a powerful technique to study the finite size effects in a solid state system. It provides a model in the real space for which the system boundaries can be easily introduced. It can also be used to study the topological aspects of the system, since it allows us to calculate boundary (edge) modes, which can determin by the topology of a band dispersion. Here, we use the six-band tight-binding model, (\ref{eq:HTBk}). Results for both zigzag and armchair edges are shown in Fig. \ref{fig:TB-ribbon-dispersion} in which the metallic and gapped edge modes are found in the zigzag and armchair ribbons, respectively.
%%%%%%
%%%%%%
In order to explore the characteristic features of zigzag and armchair ribbons, we also visualize the square of wavefunctions (SWF) for some states in both ribbons. The SWF plot provides information about the orbital character in the lattice of the system and it can be interpreted as a projected local density of states (PLDOS).
The results in Fig. \ref{fig:zig} correspond to the energy dispersion and to the SWF for the spin-up state in the zigzag ribbon where the spin-orbit coupling is taken into account. The state for each SWF plot of Fig. \ref{fig:zig} is given in the first panel of the same figure. This figure shows three edge modes where two of them cross each other within the bulk gap. This crossing edge state in zigzag MoS$_2$ has been also obtained in the three-band tight-binding model and by using density functional theory \cite{BB01,LC08,CZ14,LX13}. The figure shows that in the whole BZ of the zigzag ribbon, there are four crossings (i.e. one-dimensional Dirac cones) for the two spin and two valley flavors. We call this 1D crossing points as 1D-valleys which are different from the valleys of the bulk spectrum.  

The existence of a dispersive edge band connecting the two valleys can also be considered as the realization of a kind of "chiral anomaly", known in the quantum field theory \cite{NN83} and also in connection with Weyl semimetals \cite{HQ13}, in our 2D system. A constant electric field induces a constant flow of the electrons in momentum space, as $d {\bm k}/dt = -e{\bm  E}$. This flow can move electrons from one valley to the other one, through the edge band.  Therefore, the valley charge (or electric charge at each valley) is not conserved and this issue is called the chiral anomaly. This intervalley flow takes place, even in the absence of any source of intervalley scattering induced by broken translational symmetry. This property is related to the nontrivial topology of the system at each valley. Later on, we will discuss the topological nature of this crossing states using the simple two-band model around the K-points.

PLDOS of four particular states labeled as A, B, C and D in the first panel of Fig. \ref{fig:zig}, are shown in the bottom panels of the same figure. The states A and D correspond to the valence band maximum and conduction minimum, respectively and their orbital characters are consistent with the expected orbital character of the bulk spectrum. The bulk states of the zigzag ribbon have smooth Gaussian-like envelopes, and this form of the SWF indicates that the zigzag boundary orientation does not mix valleys. This situation is similar to the case of a graphene zigzag ribbon \cite{BF06}. Two other states ({\it i.e.} B and C) are two edge modes which are localized on opposite edges. The states labeled by B have mostly $d_{xy}$ character while the C state is mostly from $p_x$ and $p_y$ orbitals of sulfur atoms and $d_{z^2}$ orbitals of molybdenum atoms.

\begin{figure}
\centering
\includegraphics[width=0.5\linewidth]{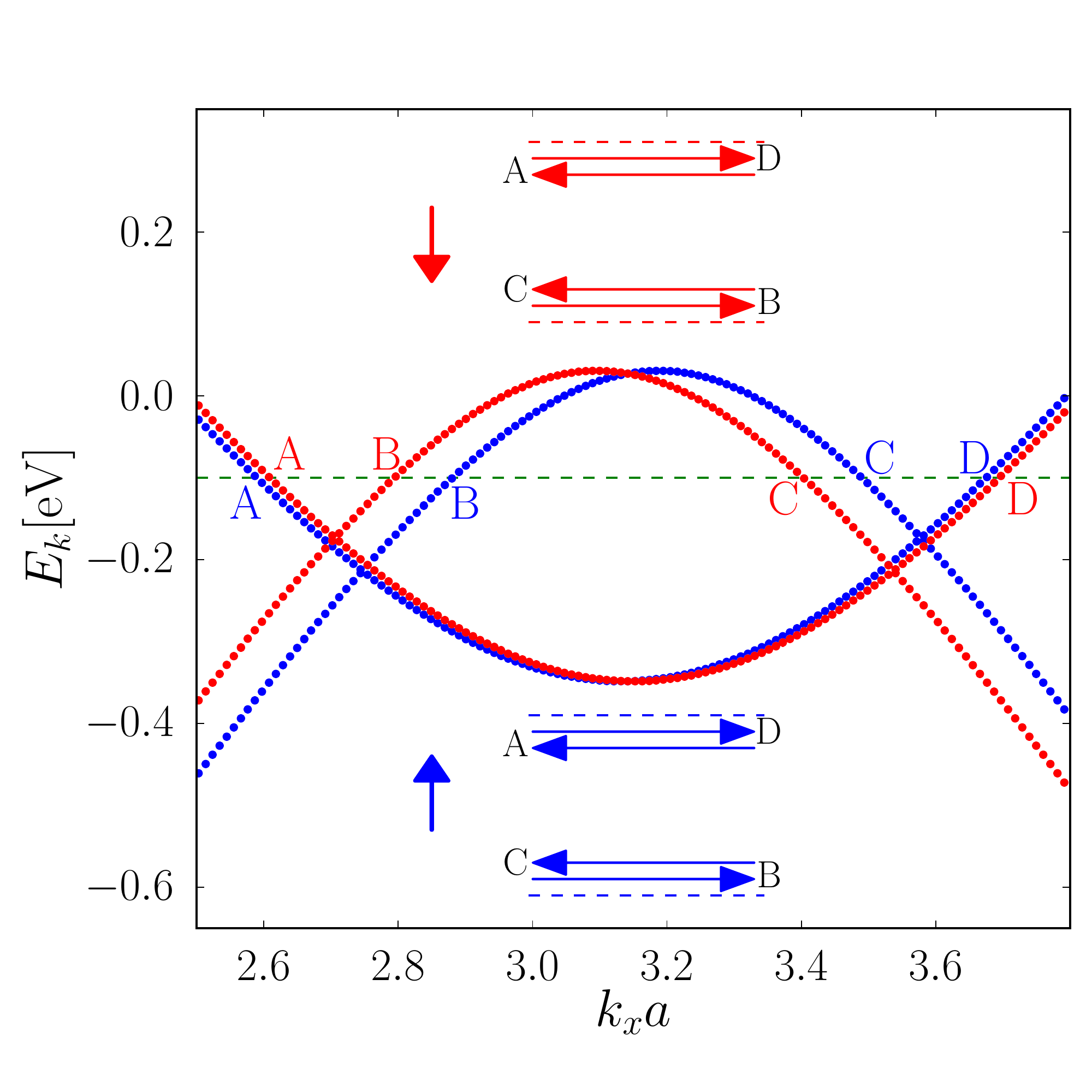}
\caption{(Color online)  Crossing edge modes in the zigzag ribbon with $N=100$. The moving direction of each edge mode is illustrated. If we add the contribution of both spin in transport channels, a quantum valley Hall (QVH) phase will be obvious in zigzag MoS$_2$.}
\label{fig:zig-edge}
\end{figure}

\begin{figure}[h!]
\centering
\includegraphics[width=0.45\linewidth]{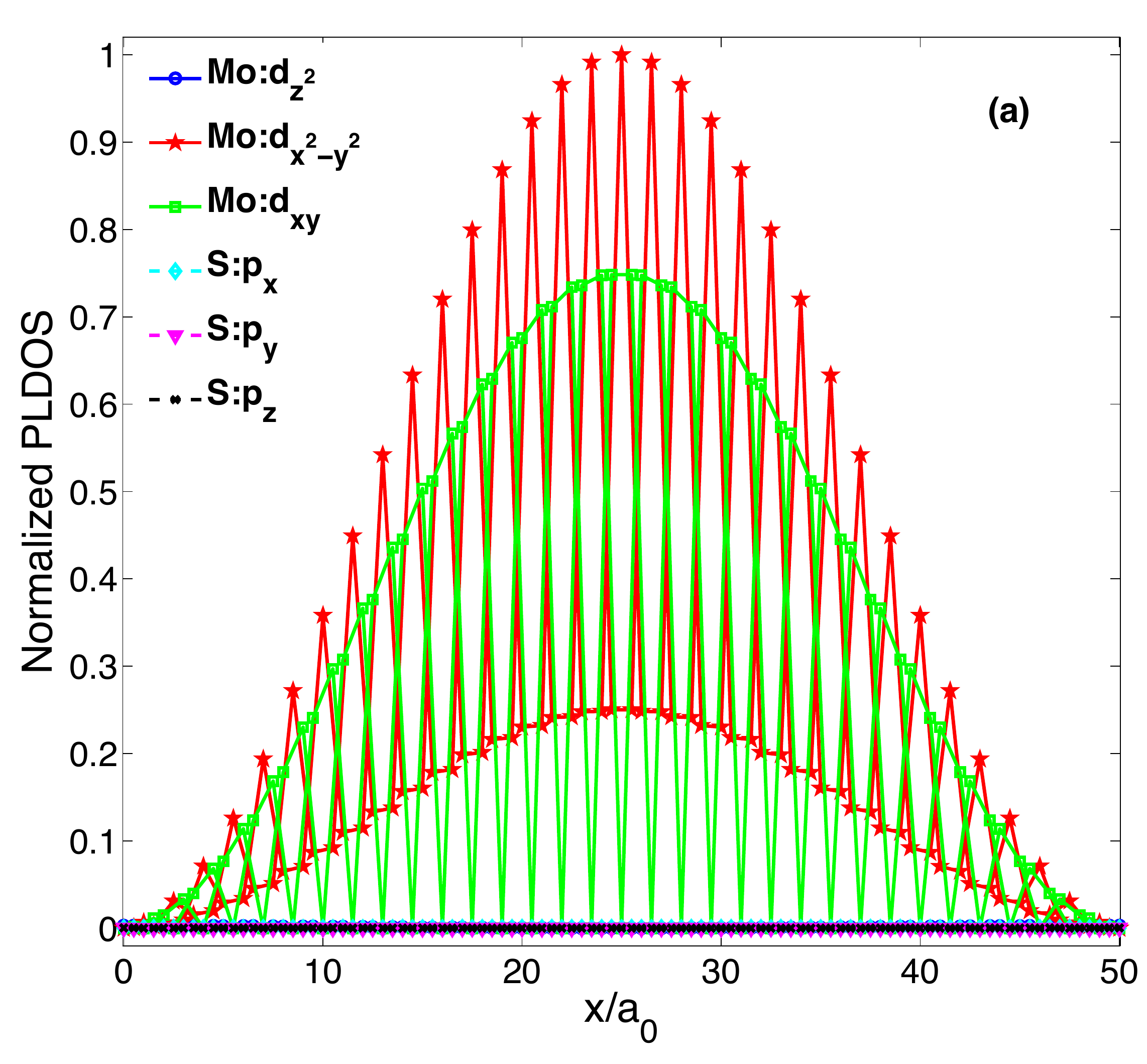}
\includegraphics[width=0.48\linewidth]{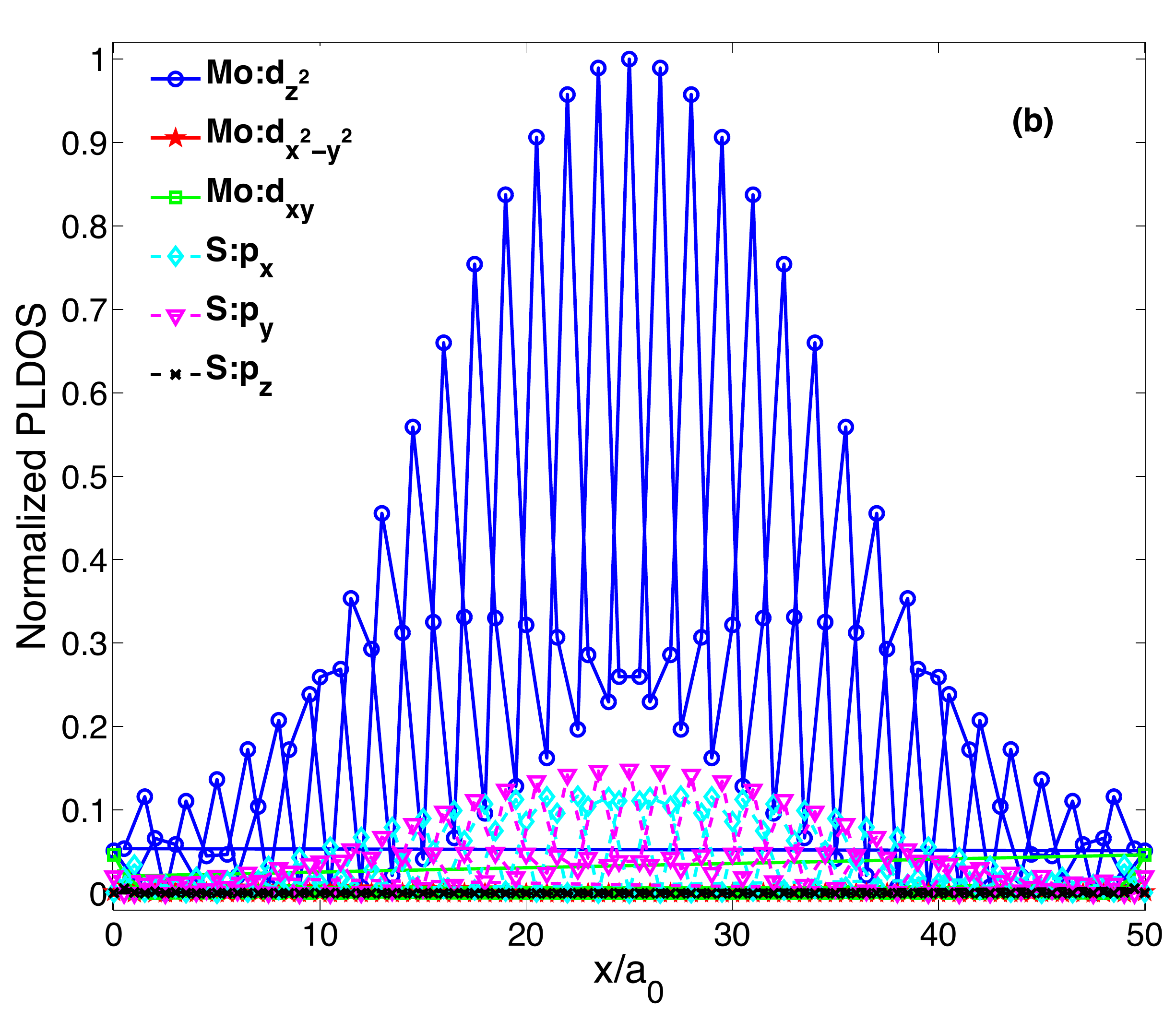}
\caption{(Color online) Projected local density of state in an armchair ribbon for the conduction band minimum and valence band maximum in panel (a) and (b), respectively. Notice that $N=101$ and the spin-orbit coupling is neglected.}
\label{fig:arm0}
\end{figure}

\begin{figure}[h!]
\centering
\includegraphics[width=0.45\linewidth]{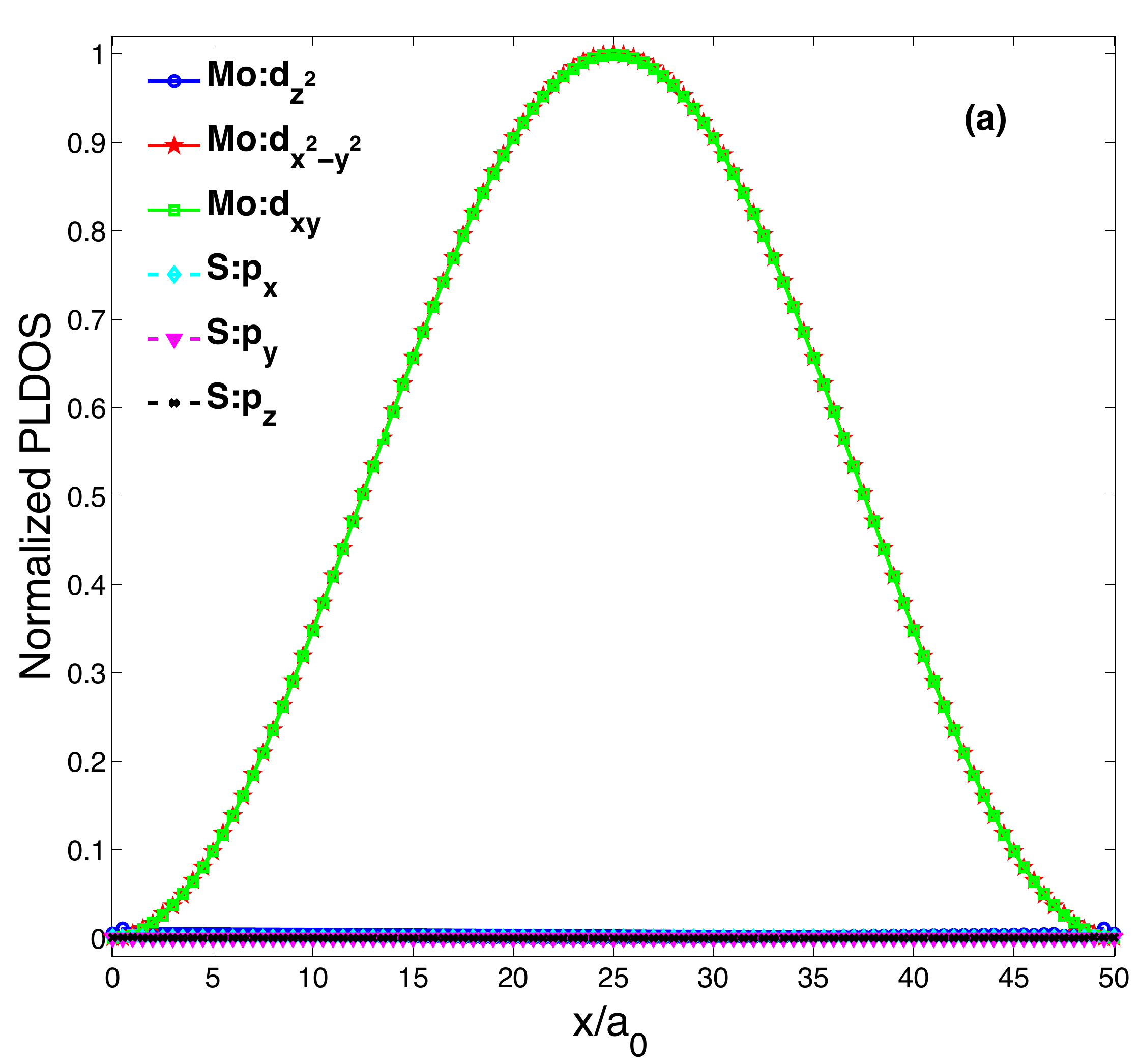}
\includegraphics[width=0.445\linewidth]{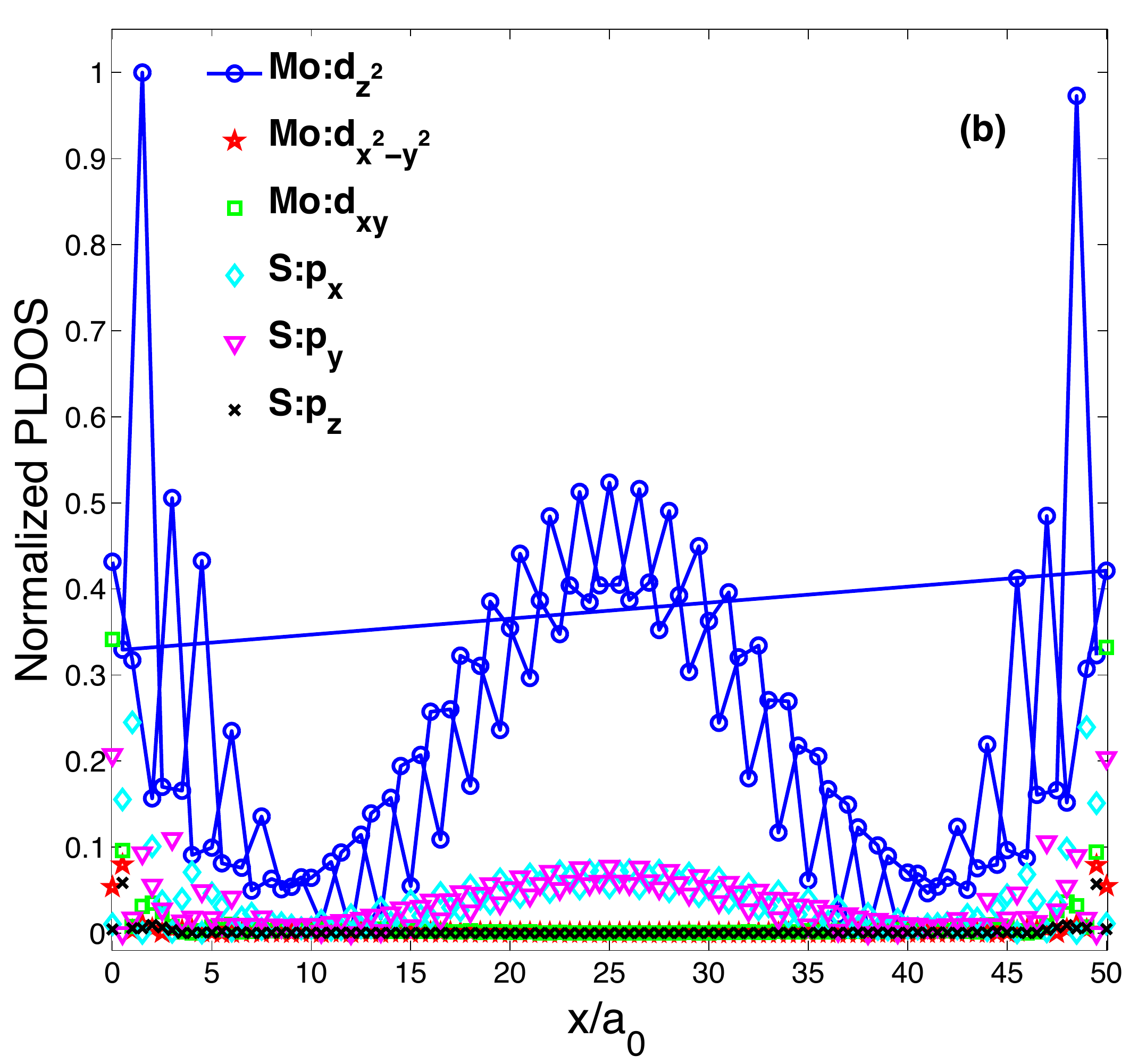}
\caption{(Color online) Projected local density of state for some states conduction band (a) and valence band (b) in an armchair ribbon with $N=101$ in the presence of the spin-orbit coupling. Two neighboring mesh points depicted by $N$ and $N+1$ labels in Fig. \ref{fig:scheam} belong to two opposite sides of the armchair ribbon. Therefore, there is a sudden jump from one side to the other side which can be seen in panel (b).}
\label{fig:arm1}
\end{figure}

\begin{figure}[h!]
\centering
\includegraphics[width=0.5\linewidth]{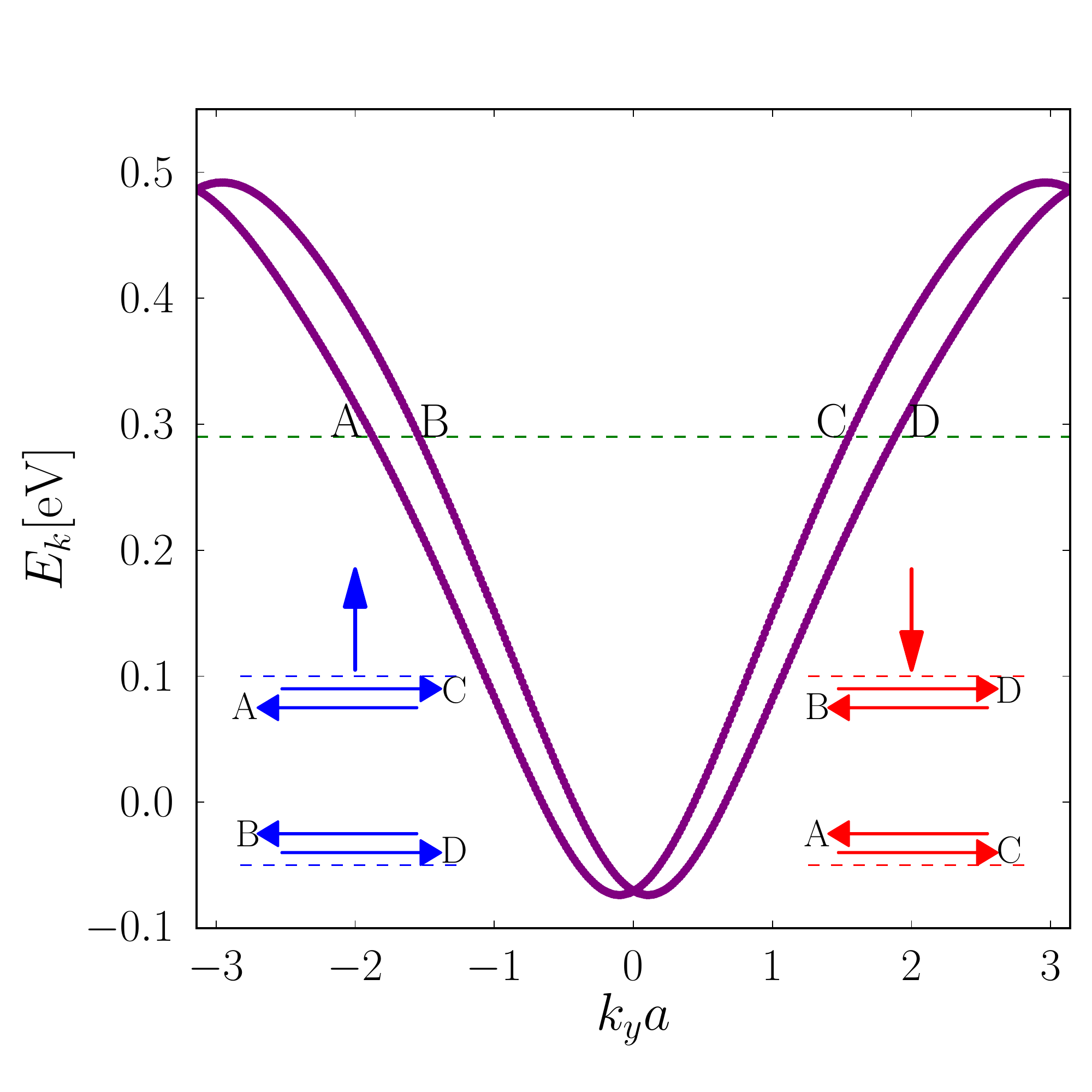}
\caption{(Color online) The chirality of the edge modes in armchair ribbon with $N=101$ which shows a pair of counter-propagating edge modes for each spin index.}
\label{fig:arm-edge}
\end{figure}

Using the SWF plots in Fig. \ref{fig:zig}, we arrive at the results depicted in Fig. \ref{fig:zig-edge} where the edge modes in two one-dimensional valleys circulate in two opposite directions leading to a weak topological insulator phase of the zigzag MoS$_2$ ribbon for each spin-flavor. Notice that in each 1D-valley, both spins circulate in the same direction, therefore this phase is actually a quantum valley Hall (QVH) state (for each spin) that is weakly protected by the time reversal symmetry in the absence of any source of a large momentum scattering.
%%%%%%
%%%%%%
%%%%%%
%%%%%%
The results in Fig. \ref{fig:arm0} correspond to the normalized projected density of states of the valence band maximum and the conduction band minimum of the armchair ribbon.
Notice the spin-orbit coupling is not included. The bulk state's wavefunctions in the valence and conduction show an oscillation, which is very similar to the expected oscillation of the wavefunctions in an armchair ribbon of graphene \cite{BF06}. The oscillation in the armchair graphene ribbon originates from a valley mixing required to satisfy the boundary condition on the edges. To understand the source of this oscillation, we consider the total wavefunction as a combination of two valleys as follows
%%%%%%
\begin{equation}
|\Psi_{\mu s}({\bm R})|^2 =|\psi_{\mu s}({\bm R})|^2
+|\psi_{\mu s}^\prime({\bm R})|^2
+ \left[\psi_{\mu s}({\bm R})\psi_{\mu s}^{\prime\ast}({\bm R})e^{i({\bm K}-{\bm K}')\cdot{\bm R}}+h.c.\right]
\end{equation}
%%%%%%
where $\mu$ and $s$ corresponds to the sublattice and spin index respectively.
If the total state is not valley-polarized, which means that both $\psi$  (wavefunction at K point) and $\psi'$ (wavefunction at K$'$ point) components are non-zero, then the square of the total wavefunction will oscillate with the period of $2\pi/|{\bm K}-{\bm K}'|$. This is the case for the armchair ribbon, see Fig. \ref{fig:arm0}, while in the zigzag case, the total wavefunction is valley-polarized and consequently there are not such oscillations.
%%%%%%
%%%%%%
The large spin-splitting in the band structure of monolayer MoS$_2$ induce a spin-valley coupling which could affect the valley mixing in the armchair ribbon. Taking into account the effect of spin-orbit coupling, we obtain the results in Fig. \ref{fig:arm1} which show the SWF of the valence and conduction in the armchair ribbon. The smooth wavefunction in the valence band originates from a suppressed valley mixing due to the large spin-splitting in the valence band. A valley-flip must occur together with a spin-flip due to the spin-valley coupling, and there is no reason for such spin-flip in a clean ribbon of monolayer MoS$_2$. A minor oscillation can be seen in the conduction band, which is related to the weakness of the spin-orbit coupling in this band. In the presence of the spin-orbit coupling, the valley mixing is suppressed, but the edge modes are still gapped. The origin of this gapped edge modes is the mixing of 1D-valleys on the edge (instead of bulk valleys) that will be discussed in the following sections.
%%%%%%
%%%%%%
The chirality of the edge modes in the armchair ribbon is summarized in Fig. \ref{fig:arm-edge} which shows a pair of counter-propagating edge modes for each spin index. From this figure, one can see that a spin-conserving backward scattering at each edge requires a large momentum transfer for edge states at high energies.
%%%%%%
%%%%%%

%%%%%%
\section{Zigzag and armchair ribbons: Low-energy model}\label{sec:KP}
%%%%%%

The armchair and zigzag ribbons have crossings in the low-energy model, which does not include valley mixing. In general, the total wavefunction can be written as a combination of two valleys, where ${\bm K}=-{\bm K}^\prime=K\hat x$, as follows
%%%%%%
\begin{equation}
\Psi_{\mu s}({\bm R})=\psi_{\mu s}({\bm R})e^{i{\bm K}\cdot{\bm R}}+\psi'_{\mu s}({\bm R}) e^{-i{\bm K}\cdot{\bm R}}~.
\end{equation}
%%%%%%
For zigzag graphene and also for monolayer $MX_2$ ribbon along x-direction, we must consider the following boundary condition
%%%%%%
\begin{gather}\label{eq:zz-bc}
 \psi_{M s}(x,y_1) e^{iK x}+ \psi'_{M s}(x,y_1) e^{-iK x}=0~,\nonumber\\
\psi_{X s }(x,y_2) e^{iK x}+ \psi'_{X s}(x,y_2) e^{-iK x}=0~.
\end{gather}
%%%%%%
This boundary condition does not mix valley indexes, since it must be fulfilled for any value of $x$ and the wavefunction at each valley should be zero, independently.
In order to satisfy above boundary condition in the case of a $MX_2$ zigzag ribbon, we use an infinite mass boundary condition by including the following potential in the Hamiltonian
%%%%%%
\begin{equation}\label{eq:im-bc}
V_{\rm hard-wall}=V_0[\theta(-y)+\theta(y-L)]
\end{equation}
%%%%%%
where $V_0 \rightarrow\infty$ and $\theta(y)$ is the step function. The infinite mass (hard-wall) boundary condition (\ref{eq:im-bc}) is different from a zigzag boundary condition (\ref{eq:zz-bc}) in general. Neither of them mix the valley indexes.
For the case of armchair ribbon along the $y$-direction, the following boundary condition must be satisfied
%%%%%%
\begin{gather}\label{eq:ac-bc}
\psi_{\mu s}(x_1,y) e^{iK x_1}+ \psi'_{\mu s}(x_1,y) e^{-iK x_1}=0~,\nonumber\\
\psi_{\mu s}(x_2,y) e^{iK x_2}+ \psi'_{\mu s}(x_2,y) e^{-iK x_2}=0~.
\end{gather}
%%%%%%
According to the above boundary condition, the valleys are mixed in an armchair ribbon. They conditions must be satisfied only for $x=x_1$ and $x=x_2$, and not for all values of $x$. Translating this valley mixing from sublattice space to band space, which is the basis of the two-band model, does not seem simple and trivial. In order to have an intuitive understanding of the edge modes in armchair ribbon, we first assume the infinite mass boundary condition given in (\ref{eq:im-bc}), which does not mix valleys then a valley mixing term, see equation (\ref{eq:inter-1D-valley}), is treated perturbatively.

%%%%%%
\subsection{Crossing edge modes in zigzag ribbon}
%%%%%%

In order to understand the crossing of edge states, we calculate the corresponding wavefunctions of these modes at $q_x$=0, using the continuum two-band model given in equation (\ref{eq:HkpK}). We also neglect the electron-hole asymmetry terms which are proportional to the identity matrix in the low energy Hamiltonian (\ref{eq:HkpK}), at $q_x$=0. Finally, the isotropic part of the Hamiltonian reads
%%%%%%
\begin{equation}\label{eq:zz_H_K}
{\cal H}^{zz}_0 =\begin{bmatrix}\Delta-\beta\partial^2_y&-v\partial_y\\ v\partial_y&-\Delta+\beta\partial^2_y\end{bmatrix}~.
\end{equation}
%%%%%%
For the sake of simplicity, we use a shortened notation as $\left (\Delta+\lambda\tau s\right)/2\rightarrow\Delta$, $\hbar^2\beta/4m_0\rightarrow\beta$ and $t_0a_0\rightarrow v$. The effect of the spin-orbit coupling on the appearance of crossing in zigzag
ribbon can be neglected. Since two spin subspaces are decoupled and the difference between these two subspaces is just a small difference on the value of the energy gap,
 we can safely drop the spin index in this section.
For the electron-hole symmetric system, the expected energy for the crossing point is $E=0$.
We calculate the eigenvalue equation at $E=0$ which leads to the following differential equation
%%%%%%
\begin{equation}
[\beta^2\partial^4_y-(2\beta\Delta+v^2)\partial^2_y+\Delta^2]\psi_c=0~.
\end{equation}
%%%%%%
The solution for this homogeneous differential equation is $e^{\xi y}$ where
%%%%%%
\begin{equation}
\beta^2\xi^4-(2\beta\Delta+v^2)\xi^2+\Delta^2=0~.
\end{equation}
%%%%%%
If $\beta\neq 0$ then one can simply find the following results for $\xi$
%%%%%%
\begin{gather}
\xi^\pm_1=\pm\frac{v-\sqrt{v^2+4\beta\Delta}}{2\beta}~,\nonumber\\
\xi^\pm_2=\pm\frac{v+\sqrt{v^2+4\beta\Delta}}{2\beta}~.
\end{gather}
%%%%%%
In the case of $\beta=0$ there are just two solutions for $\xi$ as  $\xi^\pm_0=\pm \Delta/v$.
We assume a semi-infinite system which has an edge at $y=0$ and  the general solution of the wavefunction would be as follows
%%%%%%
\begin{gather}
\psi_c(y)=A e^{\xi y}+B e^{\xi' y}~, \nonumber\\
\psi_v(y)=A \rho(\xi) e^{\xi y}+B \rho(\xi') e^{\xi' y}~.
\end{gather}
%%%%%%
in which both exponential terms must decay for $y>0$. Note that
%%%%%%
\begin{equation}
\rho(\xi) =[\frac{v}{\Delta}+\frac{\beta}{v}]\xi-\frac{\beta^2}{v\Delta}\xi^3
\end{equation}
%%%%%%
Using the hard-wall boundary condition at the edge, we have $\psi_c(0)=0$  and $\psi_v(0)=0$ which leads to
%%%%%%
\begin{gather}\label{boundary1}
A+B=0~, \nonumber\\ 
A \rho(\xi) +B \rho(\xi') =0~.
\end{gather}
%%%%%%
 A non-trivial solution of above relations satisfies $ \rho(\xi)= \rho(\xi')=\rho$, which means $\psi_v=\rho \psi_c$.
Notice that $\rho(\xi^\pm_1)=\rho(\xi^\pm_2)=\mp1$. We check different conditions in order to see when this edge state exists.
%%%%%%
\begin{itemize}
 \item $\beta=0$: There is just one decaying mode and the wavefunction cannot satisfy the boundary condition $\psi_c(0)=\psi_v(0)=0$. Hence, there is no edge state in the case of $\beta=0$.

 \item $\beta>0$: The only decaying terms are $ \xi=\xi^{+}_1$ and  $\xi'=\xi^{-}_2$, but $ \rho(\xi^{+}_1)\neq \rho(\xi^{-}_2)$. This means that there is no edge state when $\beta>0$.
  \item $\beta<0$: The only decaying terms are $ \xi=\xi^{+}_1$ and  $\xi'=\xi^{+}_2$, and also $ \rho(\xi^{+}_1)=\rho(\xi^{+}_2)=-1$  which means that there is a solution for (\ref{boundary1}) leading to the existence of edge states for $\beta<0$.
\end{itemize}
%%%%%%
The wavefunction of this edge state is given by $\psi_c=-\psi_v=\phi(y)$ where
%%%%%%
\begin{equation}
\phi(y)={\cal N} e^{ -\frac{v}{2{\mid\beta}\mid} y } \sinh[\frac{\sqrt{v^2+4\beta\Delta}}{2\beta}y]
\end{equation}
%%%%%%
in which
%%%%%%
\begin{equation}
{\cal N}=2\sqrt{\left | \frac{v\Delta}{v^2+4\beta\Delta}\right | }
\end{equation}
%%%%%%
is the normalization factor.
This state is localized at $y=0$. There are two specific (or critical) values of $\beta$ which lead to physical consequences. There is a critical value as $\beta_{cr}=-\frac{v^2}{4\Delta}$ in which the hyperbolic function is replaced by a trigonometric function. For this critical value of $\beta$ we have
%%%%%%
\begin{equation}
\psi^{cr}_c=-\psi^{cr}_v=4(\frac{\Delta}{v})^{3/2} y e^{ -\frac{2\Delta}{v} y}~.
\end{equation}
%%%%%%
Moreover, there is another critical value as $\beta'_{cr}=-\frac{v^2}{2\Delta}=2\beta_{cr}$ for which the effective mass, which depends on the second order derivative  of  the bulk energy dispersion at $q_x=0$,  changes sign. For $|\beta|>|\beta'_{cr}|$, the parabolic band changes into a Mexican hat-like dispersion.
\par
Similarly, one can show that the state localized on the other edge, {\it i.e.} $y=L$, is $\psi_c=\psi_v=\phi(L-y)$. The top (right going) and bottom (left going) edge states at the K-point are
%%%%%%
\begin{equation}\label{eq:zz-K}
|\psi_b\rangle=\begin{bmatrix}1\\-1\end{bmatrix}\phi(y) ~~~,~~~
|\psi_t\rangle=\begin{bmatrix}1\\1\end{bmatrix}\phi(L-y)~.
\end{equation}
%%%%%%
The subscript $b$($t$) stands for the state on the bottom (top) edge. Since at $q_x=0$, the velocity operator along the $x$-direction is $v_x=\sigma_x$ , we obtain $\langle v_x \rangle =\langle\psi_b|\sigma_x|\psi_b\rangle=-1$. In this case $|\psi_b\rangle$ is left-going and similarly $|\psi_t\rangle$ is right-going. These two states belong to the two branches which cross each other at $E=0,~q_x=0$. These branches are indicated by the $C$ and $D$ labels in Fig. \ref{fig:zig-edge}, shown on the bottom and top sides of the ribbon. Since, we have focused to the states exactly at the K and K$'$ points ({\it i.e.} we set $q_x=0$) the resulting Hamiltonian from (\ref{eq:zz_H_K}) is the same in both valleys. However, thanks to the time reversal symmetry we can deduce corresponding wavefunctions for the edge states at K$'$-point as follows
%%%%%%
\begin{equation}\label{eq:zz-Kp}
|\psi'_b\rangle=\begin{bmatrix}1\\1\end{bmatrix}\phi(y)~~~,~~~
|\psi'_t\rangle=\begin{bmatrix}1\\-1\end{bmatrix}\phi(L-y)~.
\end{equation}
%%%%%%
Notice that the $|\psi'_b\rangle$ ($|\psi'_t\rangle$) are located on the $B$ (A) branch shown in Fig. \ref{fig:zig-edge}.
These states in two valleys circulate in opposite directions, in order to satisfy time reversal symmetry.
The effect of trigonal warping of the position of this crossing in a zigzag ribbon can be quantified in a perturbative manner.
For a zigzag ribbon along the $x$-direction, the TW at the K and K$'$-point ($q_x=0$) is
%%%%%%
\begin{equation}
{\cal H}^{zz}_{w}=t_1\sigma_x\partial_y^2~.
\end{equation}
%%%%%%
It is worthwhile emphasizing, that the TW term at the K-point does not mix the left-going ($|\psi_b\rangle$) and the right-going ($|\psi_t\rangle$) edge states, because the spinor part of $|\psi_b\rangle$ and $|\psi_t\rangle$ are the eigenstate of $\sigma_x$ which is the spinor part of ${\cal H}^{zz}_w$. Similar features can be seen at the K$'$-point.
In other words, the TW in the zigzag ribbon commutes with the chirality operator (which is $\sigma_x$) of the one-dimensional massless Dirac edge states in the zigzag ribbon. Therefore, the gapless edge state is expected even in the presence of the TW term.\par
\begin{figure}
\centering
\includegraphics[width=0.4\linewidth]{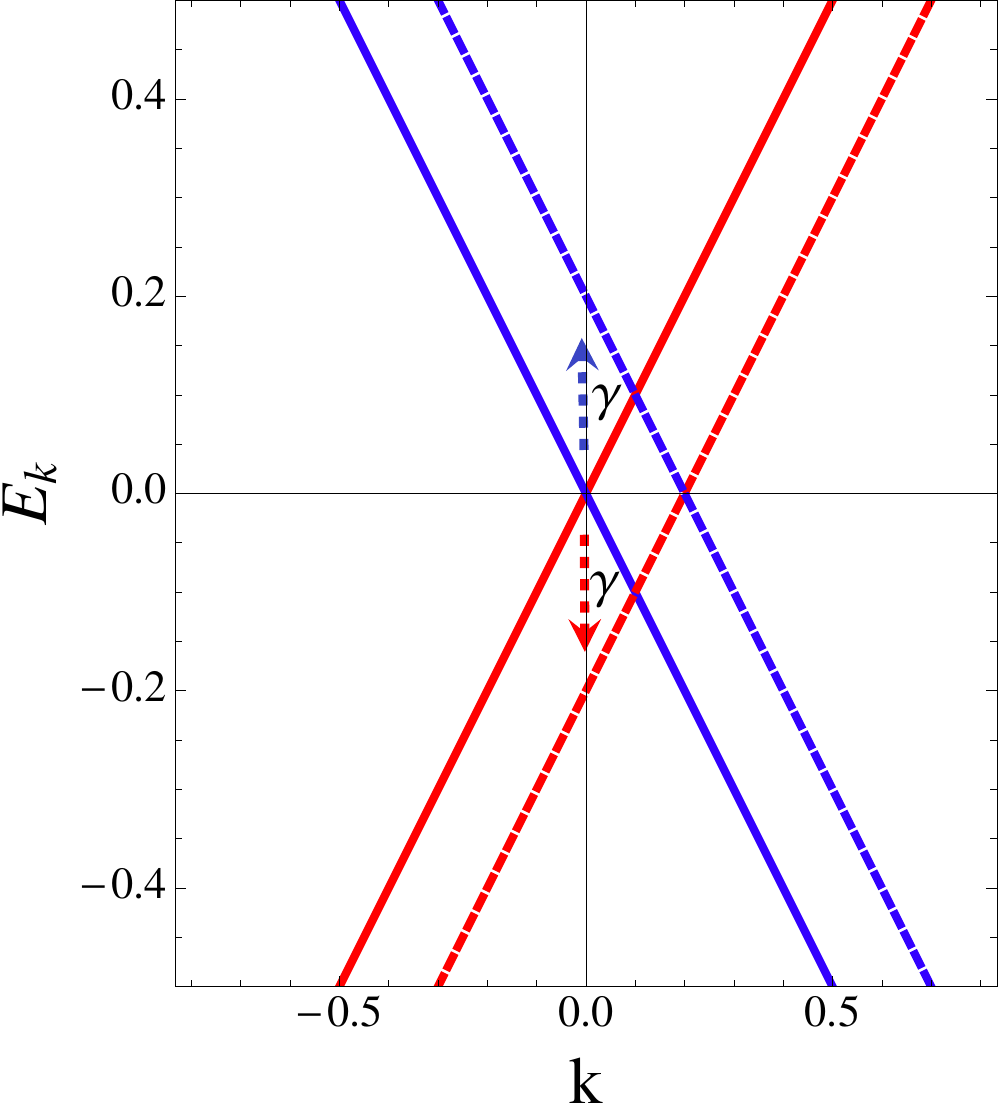}
\caption{(Color online)  The shift of the crossing point due to the TW in the zigzag ribbon. Blue (red) solid line shifts down  (up) to the blue (red) dashed line.
In this schematic figure, it is assumed that $\gamma>0$.}
\label{fig:zig-tw}
\end{figure}

We study the effect of the TW using degenerate perturbation theory. Since TW does not mix the two valleys, and also owing to the absence of the valley mixing in the zigzag ribbon, a one-valley calculation is enough for this case. Therefore, the perturbation matrix reads
%%%%%%
\begin{equation}
V^{zz}_w=\begin{bmatrix}\langle\psi_b|{\cal H}^{zz}_w|\psi_b\rangle&&\langle\psi_b|{\cal H}^{zz}_w|\psi_t\rangle\\
\langle\psi_t|{\cal H}^{zz}_w|\psi_b\rangle&&\langle\psi_t|{\cal H}^{zz}_w|\psi_t\rangle\end{bmatrix}=\gamma\sigma_z~.
\end{equation}
%%%%%%
For the case of a wide ribbon, {\it i.e.}  $L\rightarrow \infty $, we have
%%%%%%
\begin{equation}\label{eq:crossing-shift}
\gamma=\frac{t_1\Delta}{\beta} {\rm sign}\left [\frac{v^2+4\beta\Delta}{v \Delta} \right ]~.
\end{equation}
%%%%%%
This perturbation matrix means that the energy of the left going state moves up to $E_b(0)=\gamma$ while that of the right going one moves down to  $E_t(0)=-\gamma$. It seems a gap is opened at $q_x=0$, but actually the system is still gapless. This can be seen if one considers a neighborhood around $q_x=0$. If the neighborhood of $q_x=0$ moves in the same way as the states at $q_x=0$ do on each branch, one obtains
 a shift of the crossing point which is can be seen in Fig. \ref{fig:zig-tw} as a schematic visualization of the perturbation result, and also  in Fig. \ref{fig:figtw} (a) from the numerical calculations.
%%%%%%
%%%%%%
The shift of the crossing point in two valleys, owing to the TW in zigzag ribbon, must be consistent with the time reversal symmetry.

%%%%%%
\subsection{Gapped edge modes in armchair ribbon}
%%%%%%

The $y$-direction is an armchair direction, so that for a ribbon along the $y$-axis $q_y$ is a good quantum number.
For the sake of simplicity, we assume $q_y$=0. By neglecting the electron-hole asymmetric terms, the isotropic Hamiltonian at the K-point reads
%%%%%%
\begin{equation}
{\cal H}^{ac}_0 =\begin{bmatrix}\Delta-\beta\partial^2_x&-iv\partial_x\\ -iv\partial_x&-\Delta+\beta\partial^2_x\end{bmatrix}~.
\end{equation}
%%%%%%
After performing straightforward calculations, we obtain the following eigenfunction for the edge modes at $E=0$ and at the K-point
%%%%%%
\begin{equation}\label{eq:ac-K}
|\psi_b\rangle=\begin{bmatrix}1\\-i\end{bmatrix}\phi(x) ~~~,~~~
|\psi_t\rangle=\begin{bmatrix}1\\i\end{bmatrix}\phi(L-x)~.
\end{equation}
%%%%%%
Since at $q_y=0$ the velocity operator along the $y$-direction is $v_y=\sigma_y$ , so it is easy to show that $\langle v_y\rangle= \langle\psi_b|\sigma_y|\psi_b\rangle=-1$. In this case, $|\psi_b\rangle$ is a left going state and similarly $|\psi_t\rangle$ is a right going one. These two states belong to the two branches which cross each other at $E=0$ and $q_y=0$. According to (\ref{eq:ac-K}), $\sigma_y$ is the chirality operator for armchair ribbon along the $y$-direction.
To be consistent with the time reversal symmetry, we the wavefunction of the edge states at the K$'$-point is
%%%%%%
\begin{equation}\label{eq:ac-Kp}
|\psi'_b\rangle=\begin{bmatrix}1\\i\end{bmatrix}\phi(x) ~~~,~~~
|\psi'_t\rangle=\begin{bmatrix}1\\-i\end{bmatrix}\phi(L-x)~.
\end{equation}
%%%%%%
Similarly to the case of zigzag ribbon, these states in two valleys circulate in opposite directions, in order to satisfy the time reversal symmetry. The existence of this crossing is not consistent with the tight-binding results, however.
After including trigonal warping terms the crossing is not affected.  For an armchair ribbon along $y$-direction, the TW term at K and K$'$-point ($q_y=0$) is
%%%%%%
\begin{equation}
{\cal H}^{ac}_{w}=i\beta'\tau_z\otimes\sigma_z\partial_x^3-t_1\tau_0\otimes\sigma_x\partial_x^2~.
\end{equation}
%%%%%%
This trigonal warping term does not commute with the chirality operator ($\sigma_y$), which means the TW in armchair ribbon mixes the right and
the left going edge states and it can open a gap, giving rise to one-dimensional massive Dirac edge modes. The perturbation matrix in one valley, $\tau_z=1$, is
%%%%%%
\begin{equation}
V^{ac}_{w}=\begin{bmatrix}\langle\psi_b|{\cal H}^{ac}_w|\psi_b\rangle&&\langle\psi_t|{\cal H}^{ac}_w|\psi_b\rangle\\
\langle\psi_b|{\cal H}^{ac}_w|\psi_t\rangle&&\langle\psi_t|{\cal H}^{ac}_w|\psi_t\rangle\end{bmatrix}=-\gamma'\sigma_y~.
\end{equation}
%%%%%%
For $\beta<0$, for $L\rightarrow \infty$ we have $\gamma'=0$. Then, the TW itself is not enough to open a considerable gap in the edge state spectrum of armchair ribbon. The overlap of the left and right going edge states in a given valley is negligible, because they are localized on different edges. In order to open a gap we need to take into account scattering from a right to a left going state on each edge, that is, inter-valley scattering.

The existence of a gap in the armchair edge requires the hybridization of the two valleys. Since the valley mixing must change both valley and chirality of the particles, the general form of such a term must be as
%%%%%%
\begin{equation}\label{eq:inter-1D-valley}
V=(v_1\tau_x+v_2\tau_y)\otimes\sigma_x+(v_3\tau_x+v_4\tau_y)\otimes\sigma_z
\end{equation}
%%%%%%
where $\tau$ and $\sigma$ stand for the valley and band pseudospin, respectively.
To understand which of these terms are the most relevant, we must carry out a microscopic and a symmetry based analysis. We assume that all of these possible four terms are non-zero and carry out the perturbative calculation. For a wide armchair ribbon, the 1D-valley mixing potential can be written as follows
%%%%%%
\begin{equation}
V=\begin{bmatrix}
0&&0&&0&&m_{+}\\
0&&0&&m_{-}&&0\\
0&&m^\ast_{-}&&0&&0\\
m^\ast_{+}&&0&&0&&0
\end{bmatrix}
\end{equation}
%%%%%%
where $m_{\pm}=v_3\pm v_2-i(v_4\pm v_1)$ and the basis states are the following  by using (\ref{eq:ac-K}) and (\ref{eq:ac-Kp})
%%%%%%
\begin{equation}
\psi_1=\begin{bmatrix}\psi_t\\0\end{bmatrix},~
\psi_2=\begin{bmatrix}\psi_b\\0\end{bmatrix},~
\psi_3=\begin{bmatrix}0\\ \psi'_b\end{bmatrix},~
\psi_4=\begin{bmatrix}0\\ \psi'_t\end{bmatrix}~.
\end{equation}
%%%%%%
Notice that all of the overlaps between the states on opposite edges vanish exponentially and these terms have been neglected in $V$.
Performing the perturbation calculation, provides two non-zero values of the gap as $2 |m_{+}|$ and $2 |m_{-}|$.
Accordingly, 1D-valley mixing in an armchair ribbon seems to be the source of gapped edge states.
%%%%%%
\subsection{ 1D k$\cdot$p model for boundary states}
%%%%%%
We solve the eigenvalue problem in the absence of any electron-hole asymmetry terms at both the K and K$'$ points where the resulted wave functions for the zigzag case are given by (\ref{eq:zz-K}) and (\ref{eq:zz-Kp}) with zero eigenvalue. For finite $q_x$ we can deduce an effective one-dimensional ${\bm k}\cdot{\bm p}$ Hamiltonian at each valley. Following the procedure given Ref. \cite{Shen12}, we find the Hamiltonian of the edge state as
%%%%%%
\begin{equation}
{\cal H}^{zz}_{1d}=\begin{bmatrix}
\langle \psi_1|{\cal H}'|\psi_1\rangle&\langle \psi_1|{\cal H}'|\psi_2\rangle\\
\langle \psi_2|{\cal H}'|\psi_1\rangle&\langle \psi_2|{\cal H}'|\psi_2\rangle
\end{bmatrix}
=
\begin{bmatrix}
v q_x&&0&&0&&0\\
0&&-v q_x&&0&&0\\
0&&0&&-v q_x&&0\\
0&&0&&0&&v q_x
\end{bmatrix}
\end{equation}
%%%%%%
where ${\cal H}'={\cal H}(q_x)-{\cal H}(q_x=0)=\beta q_x^2\sigma_z+v \tau q_x\sigma_x$ and upper (lower) block corresponds to the K-valley ( K$'$-valley). Notice that the basis set of $\{|\psi_1\rangle,|\psi_2\rangle\}$ is equal to $\{|\psi_t\rangle,|\psi_b\rangle\}$ (\ref{eq:zz-K}) and $\{|\psi'_b\rangle,|\psi'_t\rangle\}$ (\ref{eq:zz-Kp}) for the K and K$'$ points, respectively. According to the numerical results given in Fig.\ref{fig:TB-ribbon-dispersion}, there is a small spin-splitting for the edge state spectrum in the zigzag case, but, we have neglected this splitting in the effective Hamiltonian of edge modes.

In the armchair ribbon case, we have ${\cal H}'=V+{\cal H}(q_y)-{\cal H}(q_y=0)=V+\beta q_y^2\sigma_z+v q_y\sigma_y$
where the basis set of $\{|\psi_1\rangle,|\psi_2\rangle\}$ is equal to $\{|\psi_t\rangle,|\psi_b\rangle\}$ (\ref{eq:ac-K}) and $\{|\psi'_b\rangle,|\psi'_t\rangle\}$ (\ref{eq:ac-Kp}) for K and K$'$ points, respectively. The one-dimensional model Hamiltonian of the edge modes in the armchair ribbon for the small momentum around $\rm\Gamma$ point can be obtained as follows
%%%%%%
\begin{equation}
{\cal H}^{ac}_{1d}=\begin{bmatrix}
v q_y&&0&&0&&m_{+}\\
0&&-v q_y&&m_{-}&&0\\
0&&m^\ast_{-}&&v q_y&&0\\
m^\ast_{+}&&0&&0&&-v q_y
\end{bmatrix}
\end{equation}
%%%%%%
which provides us four energy bands as follows
%%%%%%
\begin{gather}
E^{\pm}_1=\pm\sqrt{v^2 q^2_y+|m_{+}|^2}~,\nonumber\\
E^{\pm}_2=\pm\sqrt{v^2 q^2_y+|m_{-}|^2}~.
\end{gather}
%%%%%%
The energy bands given in the previous equation should be doubly degenerate at $q_y=0$ in order to be consistent with the numerical results shown in Fig.~\ref{fig:TB-ribbon-dispersion}a. Therefore, one might anticipate $|m_{+}|=|m_{-}|$ which could be possible either $v_1=v_2=0$ or $v_3=v_4=0$ is fulfilled.
In fact, the term proportional to $\sigma_x$ in (\ref{eq:inter-1D-valley}) is not consistent with the particle-hole symmetry at $\rm \Gamma$ point for each spin component. Because, it can scatter an electron-like state to a hole-like one while the electrons and holes at the  $\rm \Gamma$ point are two independent (orthogonal) eigenstates of the initial TB Hamiltonian of the armchair ribbon. Similar to the graphene case, the particle-hole symmetry operator for each spin is proportional to $\sigma_z$ \cite{AB08} and in order to respect  this symmetry we should consider the case $v_1=v_2=0$ for the boundary potential given in (\ref{eq:inter-1D-valley}) leading to  $m_{+}=m_{-}=v_3-i v_4$. 
%%%%%%
\section{Topological aspect of monolayer MoS$_2$}\label{sec:topol}
%%%%%%

\begin{figure}[h!]
\centering
\begin{overpic}[width=0.49\linewidth]{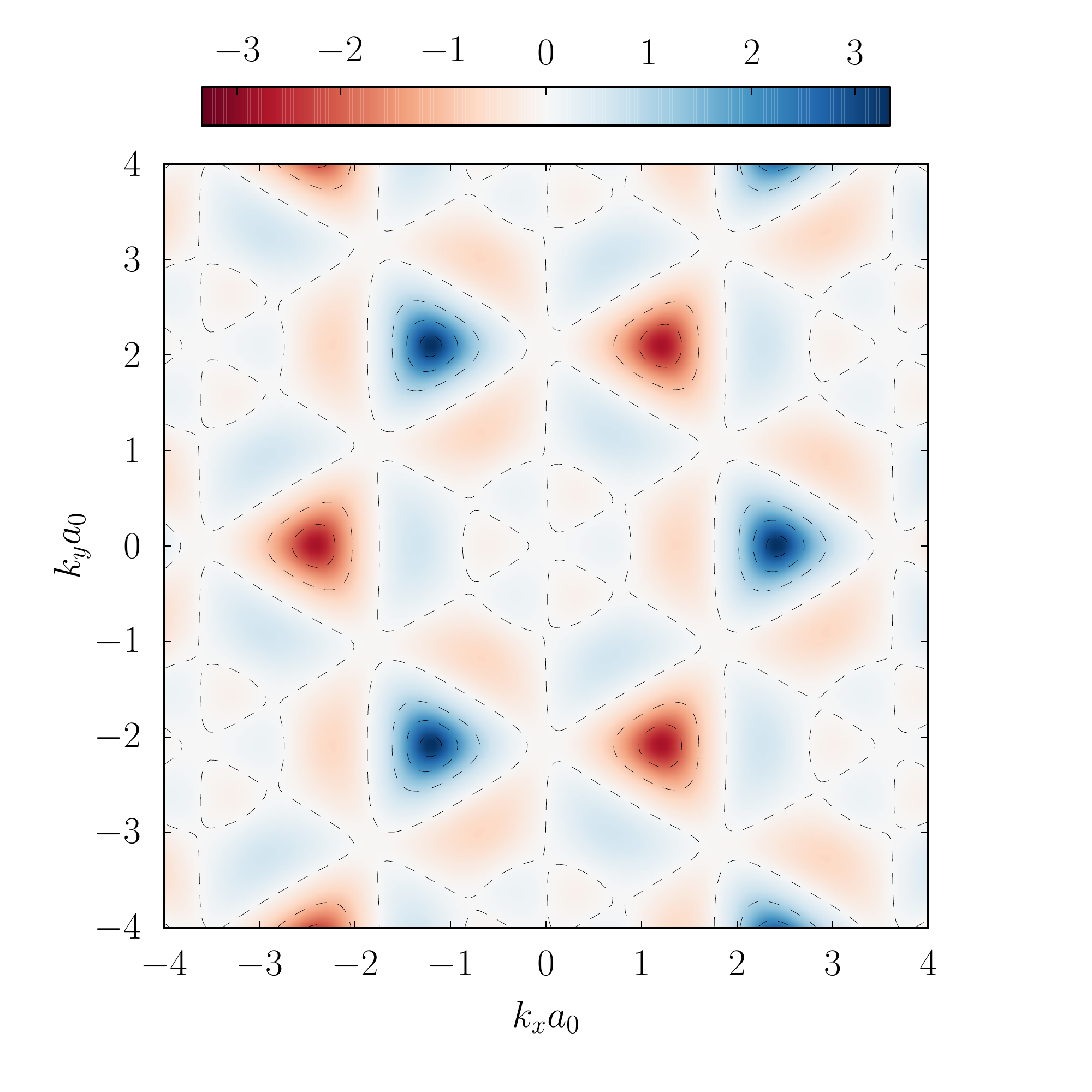}\put(0.0,85){(a)}\end{overpic}
\\
\begin{overpic}[width=0.49\linewidth]{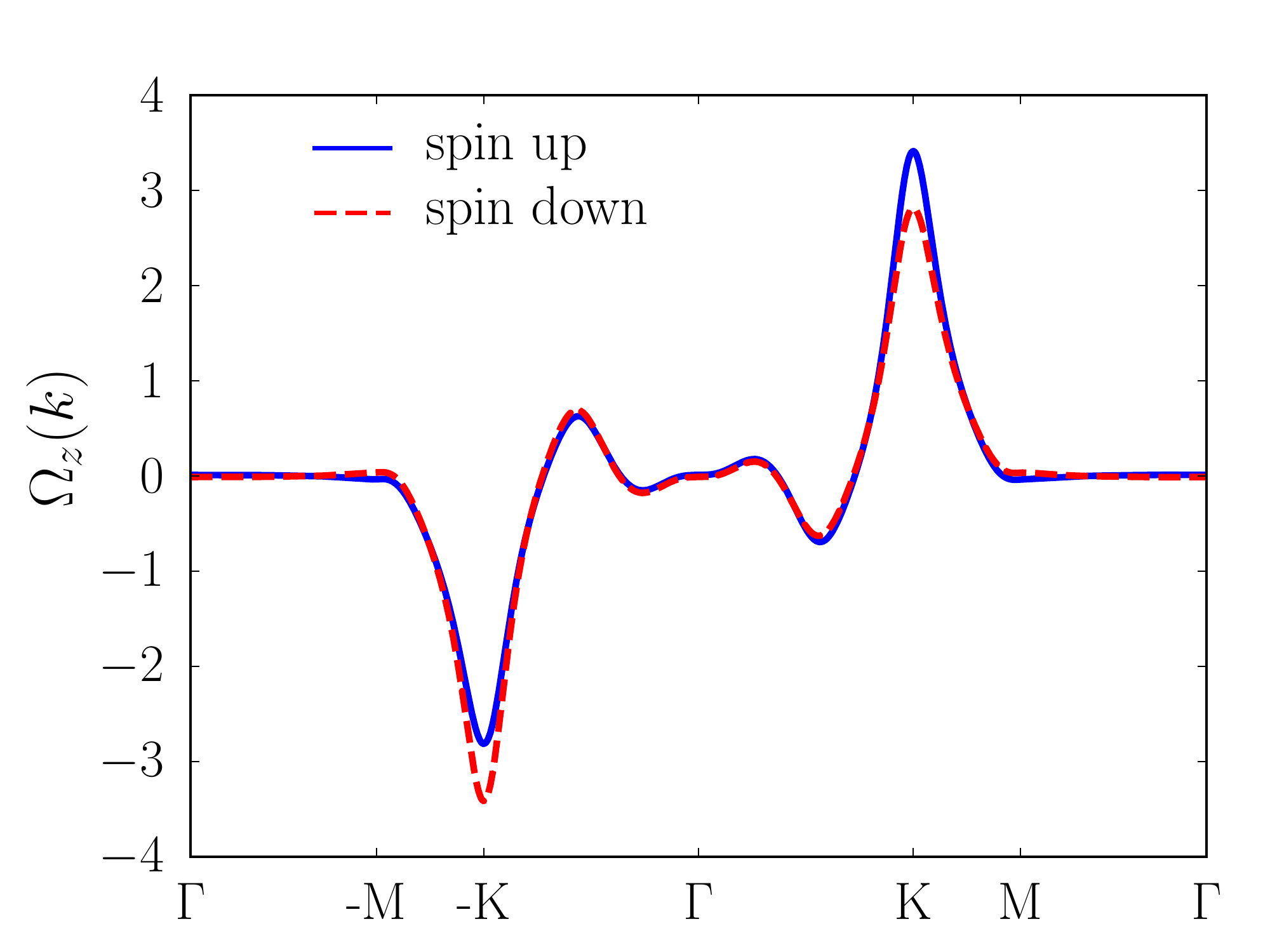}\put(85,60){(b)}\end{overpic}
\begin{overpic}[width=0.49\linewidth]{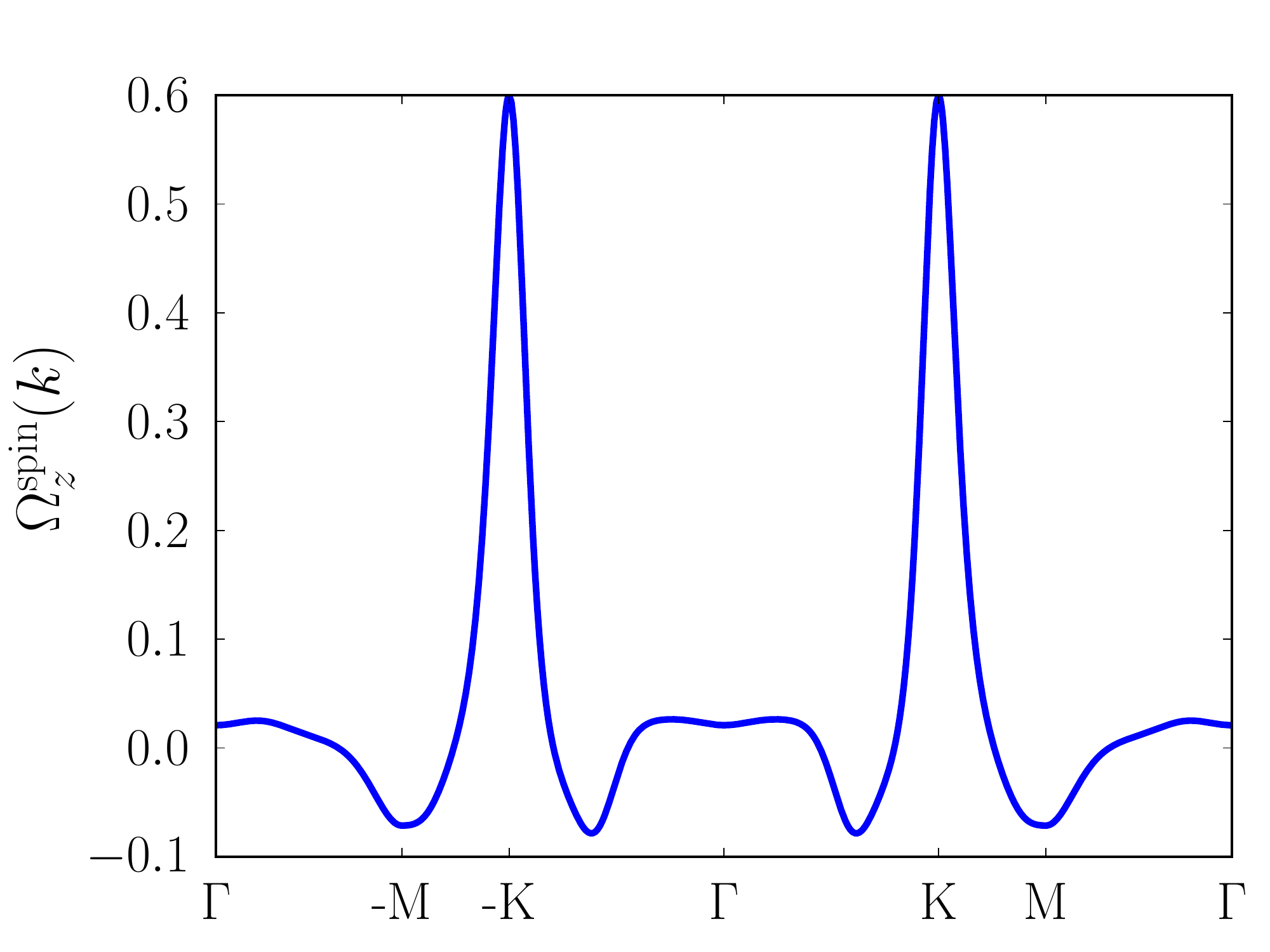}\put(87,60){(c)}\end{overpic}
\caption{ (Color online) Results Berry curvature in the whole BZ.  The panel (a) indicates the contour plot of Berry curvature (for the spin up component) in the whole BZ which clarifies hotspots around the corner of BZ. Berry curvature, panel (b), and spin-Berry curvature, panel (c), along the high symmetric point in the first BZ. Notice that the unit of Berry curvature, here, is $a^2_0$ where $a_0=a/\sqrt{3}$ with $a\approx 3.16$~\AA~as the lattice constant of the monolayer MoS$_2$.}
\label{fig:berry}
\end{figure}

The existence of the gapless edge modes in the zigzag ribbon in both tight-binding and ${\bm k}\cdot{\bm p}$ models leads us to study the topological origin of these edge modes. For this purpose, we first calculate Berry curvature in the whole BZ by using the full $k$-space Hamiltonian (\ref{eq:HTBk}). We also calculate the Chern number and the time reversal Z$_2$ invariant in the ${\bm k}\cdot{\bm p}$ model. The Berry curvature of the system is defined as
%%%%%%
\begin{equation}
\Omega_z=\sum_n \theta(\varepsilon_{\rm F}-\varepsilon_n)\Omega^n_{z}
\end{equation}
%%%%%%
in which $\Omega^n_{z}$ is the Berry curvature of the $n^{\rm th}$ band, which can be calculate through the following relation
%%%%%%
\begin{equation}
\Omega^n_{z}=i\sum_{n'\neq n} \frac{\langle n|\hbar v_x|n'\rangle\langle n'|\hbar v_y|n\rangle-(x\longleftrightarrow y)}{(\varepsilon_n-\varepsilon_n')^2}
\end{equation}
%%%%%%
where $\varepsilon_n$ and $|n\rangle$ stands for the dispersion and wave vector for $n^{\rm th}$ band.
 The Fermi energy ($\varepsilon_{\rm F}=0$) lies inside the main gap and $\theta(x)$ is the step-function which stands for the Fermi-Dirac distribution function. The velocity operator is defined as $\hbar v_i=\partial {\cal H}^{\rm TB}_{s}({\bm k})/\partial k_i$.
In  Fig. \ref{fig:berry}, the contour plot of the Berry curvature for the spin up component is shown. The results from six-band TB calculation are very similar to those obtained using DFT \cite{FX12}. The calculated Berry curvature using the six-band model shows that the states around the K-point give the main contribution to the topological Berry curvature. The Berry curvature has opposite signs in the two valleys, in order to satisfy time reversal symmetry.
In Fig. \ref{fig:berry}, we plot the Berry curvature along the high symmetry directions in the first BZ. A large value of the Berry curvature can be seen near to the K and K$'$ points. The right panel in Fig. \ref{fig:berry} indicate spin Berry curvature ($\Omega^{\rm spin}_z=\Omega^\uparrow-\Omega^\downarrow$).
This non-zero value for the spin Berry curvature leads to the existence of an intrinsic spin Hall conductivity in monolayer MoS$_2$ that has been also studied through {\it ab-initio} methods \cite{FX12}.
%%%%%%
%%%%%%

The existence of larger values of the Berry curvature around the K-point as compared to the other regions in the BZ indicates that the low-energy isotropic Hamiltonian around the K-point is enough to study qualitatively the topological aspects of the monolayer MoS$_2$ band structure. We now consider the topological structure of the low energy model around the K points, equation (\ref{eq:HkpK}).

\begin{figure}
\centering
\includegraphics[width=0.5\linewidth]{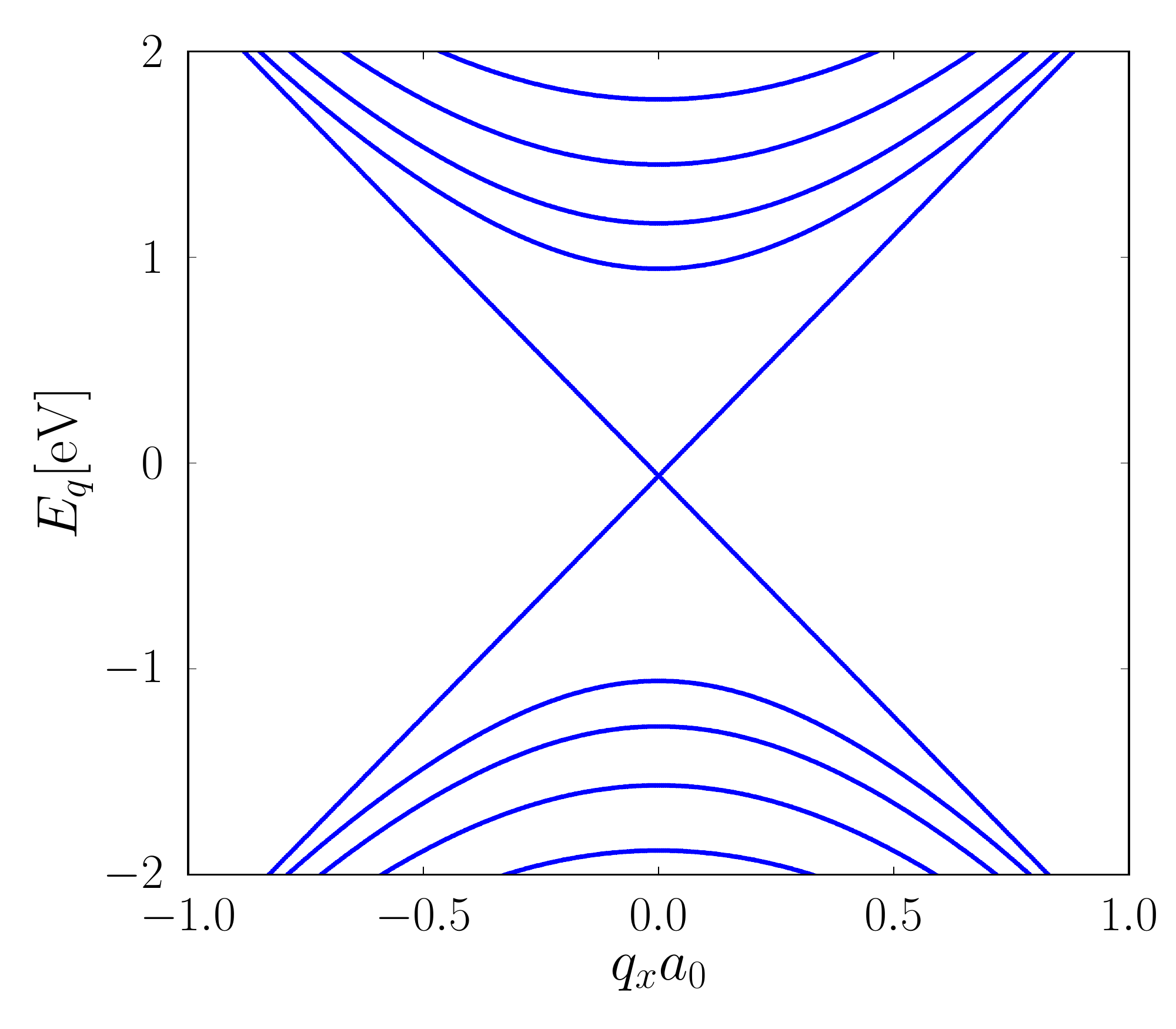}
\caption{(Color online) Crossing edge modes from the isotropic model Hamiltonian around K-point for spin up, ${\cal H}^i_{\tau=+,s=+}$, as given in (\ref{eq:HkpK}). There is a small shift of crossing point from $E=0$ which is due to the presence of $\alpha$ in our numerical calculation which breaks the particle-hole symmetry.}
\label{fig:crossing2b}
\end{figure}

The isotropic part of the two-band Hamiltonian, ${\cal H}^i_{\tau s}$, is a modified-Dirac Hamiltonian containing a momentum dependent mass term \cite{Shen12,BHZ06}. This diagonal quadratic term, proportional to $\beta$, in the Hamiltonian has a topological meaning. We calculate the Chern number of this Hamiltonian at each valley, following the well-known relation for a Hamiltonian as ${\cal H}=\epsilon(k)+{\bm \sigma}.{\bm d}$ in which ${\bm \sigma}=\sigma_x\hat{x}+\sigma_y\hat{y}+\sigma_z\hat{z}$. Then, the Chern number reads
%%%%%%
\begin{equation}
C=\frac{1}{4\pi}\int_{-\infty}^{\infty}\int_{-\infty}^{\infty}\frac{(\partial_{q_x}{\bm d}\times\partial_{q_y}{\bm d})\cdot {\bm d}}{|{\bm d}|^3}dq_x dq_y~.
\end{equation}
%%%%%%
Notice that $\epsilon(k)$, which is the effective mass asymmetry term (i.e. $\alpha$, $\lambda_0$ and $\Delta_0$), has no contribution to the Chern number.
The Chern number from ${\cal H}^i_{\tau s}$ at the K-point for each spin is
%%%%%%
\begin{equation}
C_{Ks}=\frac{1}{2}(\rm sign(\Delta+\lambda s)-\rm sign(\beta))
\end{equation}
%%%%%%
where the Chern number of the K-point is $C_{K}=C_{K\uparrow}+C_{K\downarrow}$ and for the other valley, we have $C_{K'}=-C_{K}$. The total Chern number ($C=C_{K}+C_{K'}$) is zero which is consistent with time reversal symmetry. From the negative numerical value of the $\beta$, it follows that ${\cal H}^i_{\tau s}$ has non-trivial topology if we consider one valley and take the spin as a good quantum number. This non-trivial topology results in a crossing of the edge modes, which can be seen in Fig. \ref{fig:crossing2b} to satisfy the bulk-edge correspondence. To obtain this figure, we have used a finite element discretization of the low-energy Hamiltonian which has been discussed in Ref. \cite{RG15}.
%%%%%%
%%%%%%
Since there are four crossing points (2 for the spin and 2 for the valley), this state of the system is not strongly topological. A local perturbation might be able to hybridize the edge modes with opposite topology and open a gap at the crossings. In other words, since the number of gapless edge modes is even (here it is 4) the system can be classified as a weak topological insulator. This weak topological structure (for each spin) will be destroyed, if two valleys mix with each other owing to any valley mixing caused by edge orientation, edge roughness, and short-range scatterers like sulfur vacancies in the monolayer MoS$_2$. A non-zero valley Chern number as $C_v=C_{K}-C_{K'}$ can be defined that indicates a quantum valley Hall phase for each spin. The topology of the band structure will never changes if the energy gap in the band structure does not close and reopen as the model parameters vary continuously. To understand this topological phase, one can assume $\lambda=0$ and tune $\Delta$ from positive to a negative value, so that there is topological phase transition from a non-trivial to a trivial state in the case of $\beta<0$.

According to the low-energy model, there is a non-zero valley Chern number, while the total Chern number is zero based on the time reversal symmetry ($\Theta$). Since the total Chern number is zero, it is required to explore another topological invariant which is the time reversal Z$_2$ invariant. In the following, we calculate this invariant through the well-known Pfaffian approach introduced by Kane and Mele \cite{KM05}. The Pfaffian of the time reversal operator is indicated as $P({\bm q})={\rm Pf}[\langle \psi_i ({\bm q})|\Theta|\psi_j({\bm q})\rangle]$ where $\Theta$ and $\psi_j({\bm q})$ are the time reversal symmetry operator and the wavefunction of $j^{\rm th}$ state, respectively. In {\ref{app:z2}}, the calculation for the Pfaffian of time reversal operator is given in detail. For the isotropic low-energy model, we find $P({\bm q})=\left [P_{+}({\bm q})P_{-}({\bm q})\right ]^2$ in which
%%%%%%
\begin{equation}
P_{\pm}({\bm q})=\frac{\frac{\Delta \pm\lambda}{2}+b\beta |a_0{\bm q}|^2}{\sqrt{(\frac{\Delta \pm\lambda}{2}+b\beta |a_0{\bm q}|^2)^2+t_0^2 |a_0{\bm q}|^2}}~.
\end{equation}
%%%%%%
We remind that $b=\hbar^2/(4m_0 a^2_0)$. Since $(\Delta\pm\lambda)/\beta<0$, there are zeros for this Pfaffian on two circles with $|a_0 {\bm q}_{\pm}|^2=(\Delta\pm\lambda)/(2b|\beta|)$ as their radii. According to the Kane-Mele prescription\cite{KM05}, the time reversal Z$_2$ topological invariant can be calculated as follows
%%%%%%
\begin{equation}
\nu=\frac{1}{2\pi i}\int_{\cal C} d{\bm q}\cdot{\bm \nabla} \log[P({\bm q})+i\delta]
\end{equation}
%%%%%%
where $\delta$ is an infinitesimal real positive number and ${\cal C}$ is a contour surrounding the half of BZ.
Following the procedure given in Ref. \cite{SS10}, the contour of ${\cal C}$ is the upper half-plane and the integral on the arc is zero since $\delta>0$ and the arc has an infinite radius. Therefore, only the integral on the $q_x$ axis remains, which contains four poles. Since $\delta>0$ only two of them contribute in the Z$_2$ invariant. Then, $\nu=2\times\frac{1}{2\pi i}\times [2\pi i+2\pi i]~\text{Mod}~2=0$ which shows the trivial nature of the system. If we only consider one spin and one valley ({\it i.e.} spinless and valleyless system ), the $\Theta$ matrix will be a $2\times2$ matrix that provides $\nu=1$ indicating the non-trivial nature of the modified Dirac model for each spin and valley.

\begin{figure}[h!]
\centering
\includegraphics[width=0.32\linewidth]{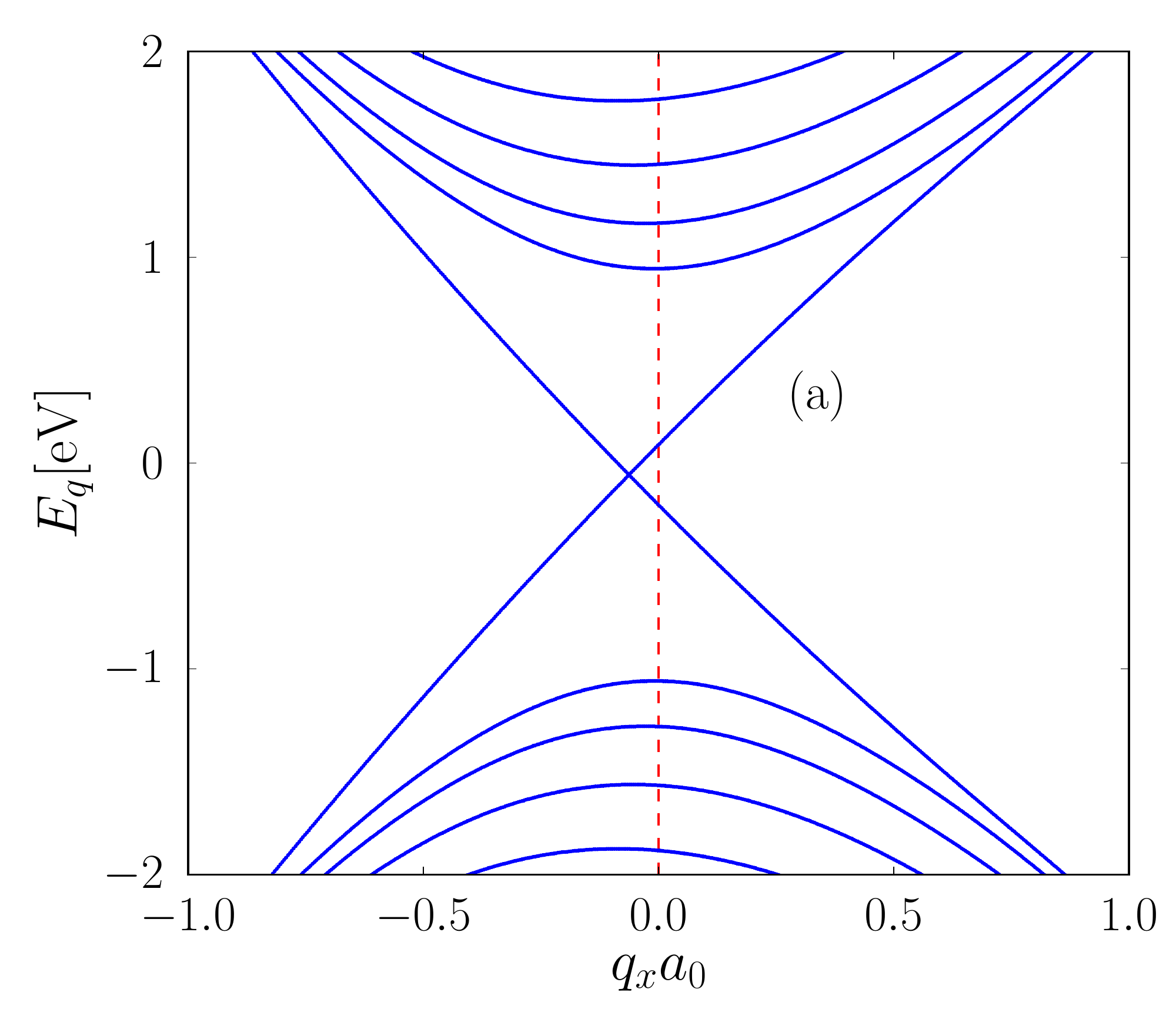}
\includegraphics[width=0.32\linewidth]{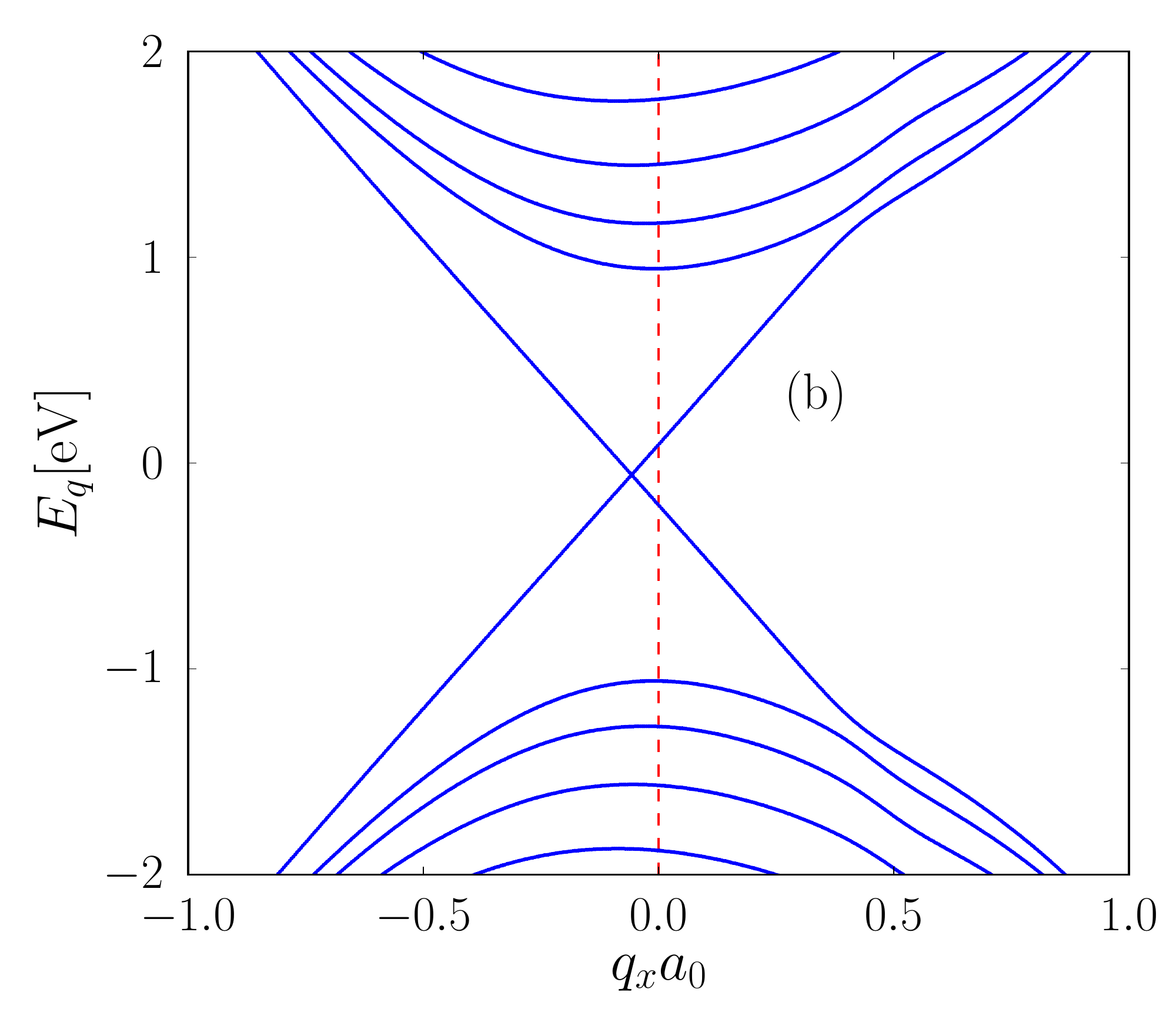}
\includegraphics[width=0.32\linewidth]{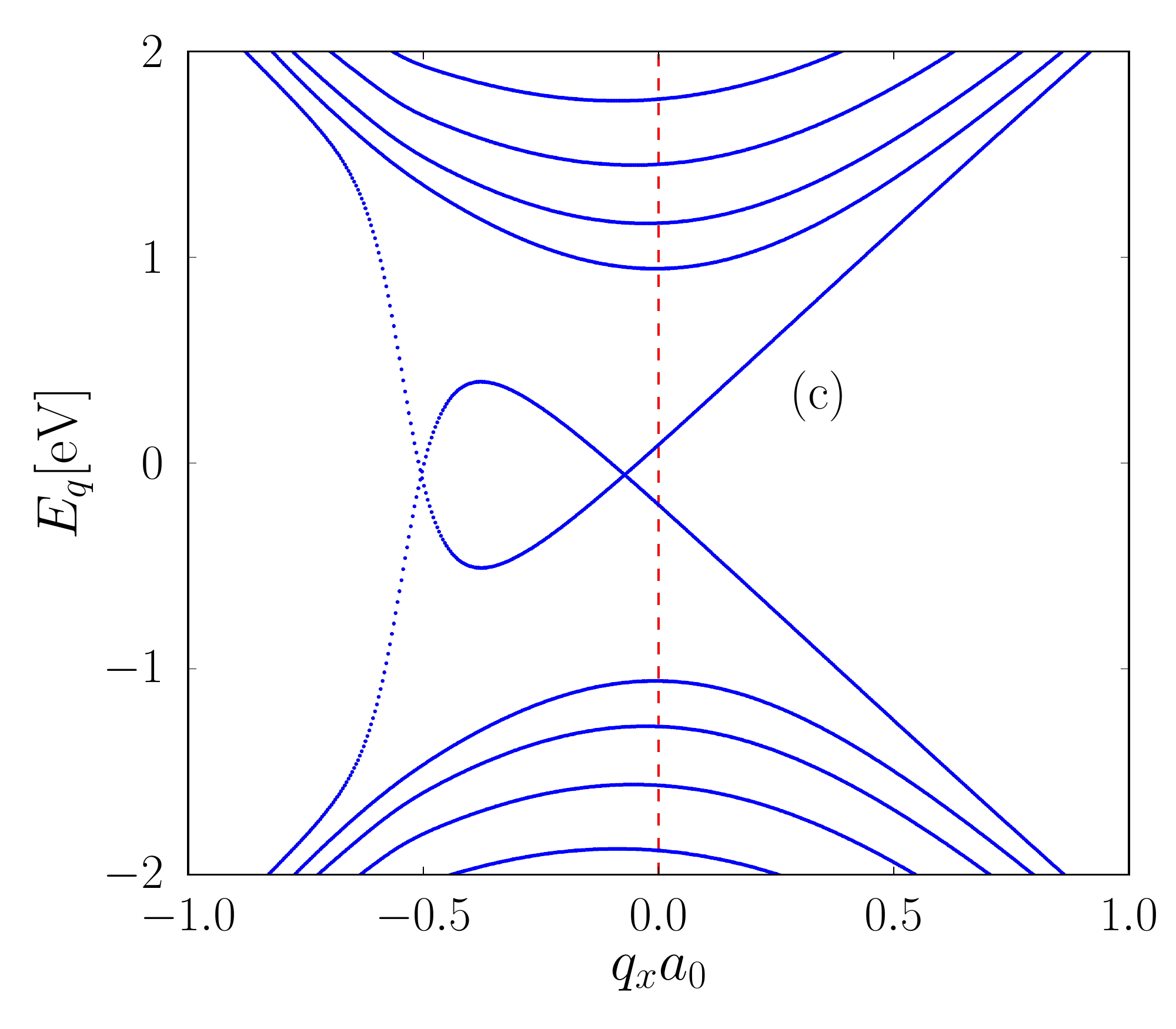}
\caption{(Color online) Energy dispersion obtained from low-energy model for a ribbon along $x$-direction for the case of K valley ($\tau=+$) and spin up ($s=+$). (a) $t_1=-0.144$~eV and $\beta'=0$. (b)  $t_1=-0.144$~eV and $\beta'=-0.534$  which is based on our low energy model of MoS$_2$. (c)  $t_1=-0.144$~eV and $\beta'=+0.534$. The crossing point shifts away from the K-point as a result of the trigonal warping. Moreover, the number of crossing point could change for different sign of $t_1\times \beta'$. }
\label{fig:figtw}
\end{figure}

In the low-energy model, there is also a trigonal warping term which might have a contribution to the band topology. In the following, we investigate this contribution by using
the trigonal warping term of low-energy Hamiltonian, ${\cal H}^w_{\tau s}$, as given in (\ref{eq:HkpK}). Notice that we neglect the term proportional to $\alpha'$ in this survey, since it is a trivial term similar to the term proportional to $\alpha$. The calculation of the energy dispersion for a ribbon in the absence of the TW, see Fig. \ref{fig:crossing2b}, leads to a pair of edge states for each spin and valley index which closes the gap.  In principal, the trigonal warping could have some topological consequences on the edge mode dispersion. After including trigonal warping in the model, the odd number of the crossing points for each flavor, which is one, can increase or decrease to an even number, depending on the sign for $t_1\times \beta'$. This possible change in the number of crossing points implies a kind of topological phase transition. This feature is evident in the numerical results shown in Fig. \ref{fig:figtw}. It should be noticed that the results in Fig. \ref{fig:figtw} correspond to the zigzag ribbon along $x$-direction which has a finite width in the $y$-direction. Therefore, in ${\cal H}^w_{\tau=+, s=+}$, we implement $q_y \equiv -i\partial_y$ and $q_x$ as a good quantum number.

In Fig. \ref{fig:figtw} (a) we just consider the term proportional to $t_1$ in the TW low-energy model and a shift of the crossing point along the horizontal axis ($q_x$) is evident in this figure.  In the anisotropic case, the crossing does not occur exactly at the K-point and this is consistent with our TB results shown in Fig. \ref{fig:TB-ribbon-dispersion} and also with our perturbative analytical results shown in equation (\ref{eq:crossing-shift}). The shift of crossing point in the TB results shown in Fig. \ref{fig:TB-ribbon-dispersion} is larger than that of the low-energy model which is depicted in Fig. \ref{fig:figtw} (a). The reason for this mismatch is related to the limited validity range of the low-energy model for large $|{\bm q}|$.
In the panel (b) and (c) of this figure, the term of $\beta'$ in the TW model Hamiltonian is also turned on. According to the panel (b), we conclude that the inclusion of the TW in the low-energy model of monolayer MoS$_2$ does not change the topology of the band structure, which is consistent with our full TB calculations. However,  as it is obvious in the panel (c) of the same figure, the number of crossing  changes from one to two when the sign of $t_1\times \beta'$ is changed from positive to negative.
%%%%%%
%%%%%%

Finally, it is worth to mention that in a finite size calculation of a polar material with non-zero polarity of the bulk bands, it is expected to have a charge rearrangement on the edge owing to the Coulomb interaction \cite{GC13,GM15} which is not considered in our model. This rearrangement of charge can be generally modeled in a self-consistent TB manner. In order to take that into account, first of all, we need to estimate the change in the on-site energy of the atomic site on the edge and then to implement this change in the TB calculation in a self-consistency way to achieve a convergence in the edge modes dispersion. 

In our system, we have estimated the change of the on-site energy as $ \delta E \sim \delta P \times U$ with respect to its value in the bulk where $U\sim 1$~eV is a typical value for the Hubbard interaction on the same site and $\delta P = P_{\rm bulk}- P_{\rm edge}$ in which $P_{\rm edge}$ ($P_{\rm bulk}$) stands for the probability of finding an electron in the d-orbitals of Mo atoms on the Mo-terminated edge of the zigzag ribbon (the bulk) of monolayer MoS$_2$. According to our six-band TB model, we have $\delta P \sim 0.89-0.79 =0.1$. The change in the on-site energy can shift the edge bands by $w\times \delta E$ where $w = |\psi_{\rm edge} (y=0)|^2$ is the weight of the states in the band localized on the edge atoms. Based on our TB model, we can take $w \sim 0.1-0.5$ as a typical value of the weight on the Mo-terminated edge of the zigzag ribbon.Therefore, we conclude that $ w\times  \delta P \times U\sim 0.01-0.05 $ eV as the shift of the edge mode energy due to the charge rearrangement which is too small and it can not change edge modes dispersion in a wide ribbon. Accordingly, we believe a static TB calculation would be enough to capture main electronic feature of MoS$_2$ ribbon with a wide width e.g. $N>100$.

%%%%%%
\section{Conclusion}\label{sec:conclusion}
%%%%%%

In this work, we have studied the finite size effect in the monolayer MoS$_2$ nanoribbon as a representative of monolayer TMDs.
We consider a tight-binding model which is invariant with respect to the horizontal reflection symmetry of the system and contains
three $d$ and three $p$ orbitals describing the Mo and S atoms, respectively. By employing this microscopic tight-binding model,
the band structure of both zigzag and armchair ribbons are obtained. We investigate the nature of the edge modes for these
two edge terminations, and our numerical results reveal metallic and gapped edge modes for the zigzag and armchair case, respectively.
The metallic (gapped) nature of the edge modes in the zigzag (armchair) ribbon is consistent with {\it ab-initio} calculation and experimental results.
In contrast to the case of an armchair graphene ribbon, there is no valley mixing in the valence band of monolayer MoS$_2$ owing to the strong spin-orbit coupling in this system.
\par
More intuitive analytical calculations by considering a low-energy ${\bm k}\cdot {\bm p}$ clarifies a weak topological protection for the metallic edge modes in the zigzag ribbon. The topological nature of monolayer MoS$_2$ originates from the particular orbital character of each energy band. Owing to the large value of the energy gap in comparison with the spin-orbit coupling, a topological phase transition is not possible by including (or neglecting) the effect of the spin-orbit coupling.
This calculation implies that a crossing of edge modes is expected also for boundaries induced by external gating. We have shown that the crossing point of the edge modes in the zigzag ribbon is not located on the K point and it shifts away from the K point due to the effect of trigonal warping. We have emphasized a one-by-one correspondence between the low-energy model parameters and edge mode dispersion, particularly in topological picture. 
We have finally shown that the mixing of edge modes in two different 1D-valleys results in a gapped spectrum of the edge modes in the armchair
ribbon of monolayer MoS$_2$.

%%%%%%
\section{Acknowledgements}
%%%%%%

We acknowledge funding from the European Commission under the Graphene Flagship, contract CNECTICT- 604391. F. G. is partially funded by ERC, grant 290846, and MINECO (Spain), grant FIS2014-57432. H. R. was supported by Fondazione Istituto Italiano di Tecnologia and the European Union's Horizon 2020 research and innovation programme under Grant Agreement No. 696656 GrapheneCore1. R. A. and H. R. acknowledge for partially support from Iran Science Elites Federation under Grant No. 11/66332. H. R.  thank Pablo San-Jose and Rafael Rold\'an for very useful discussions and technical help. 
%%%%%
\appendix
\section{Pfaffian of the time reversal operator}\label{app:z2}
%%%%%%

To calculate the Pfaffian of the time reversal operator, we first construct the matrix representation of the time reversal symmetry in a space with 8 basis functions as follows
%%%%%%
\begin{eqnarray}
&&\psi^c_{K\uparrow}(q)=\frac{1}{D^c_{+}}\begin{bmatrix}-t_0 q^\ast\\h^c_{+}\\O\\O\\O\end{bmatrix}
~~~,~~~\psi^v_{K\uparrow}(q)=\frac{1}{D^v_{+}}\begin{bmatrix}-t_0 q^\ast\\h^v_{+}\\O\\O\\O\end{bmatrix}
\nonumber\\
&&\psi^c_{K\downarrow}(q)=\frac{1}{D^c_{-}}\begin{bmatrix}O\\-t_0 q^\ast\\h^c_{-}\\O\\O\end{bmatrix}
~~~,~~~\psi^v_{K\downarrow}(q)=\frac{1}{D^v_{-}}\begin{bmatrix}O\\-t_0 q^\ast\\h^v_{-}\\O\\O\end{bmatrix}
\nonumber\\
&&\psi^c_{K'\uparrow}(q)=\frac{1}{D^c_{-}}\begin{bmatrix}O\\O\\t_0 q\\h^c_{-}\\O\end{bmatrix}
~~~,~~~\psi^v_{K'\uparrow}(q)=\frac{1}{D^v_{-}}\begin{bmatrix}O\\O\\t_0 q\\h^v_{-}\\O\end{bmatrix}
\nonumber\\
&&\psi^c_{K'\downarrow}(q)=\frac{1}{D^c_{+}}\begin{bmatrix}O\\O\\O\\t_0 q\\h^c_{+}\end{bmatrix}
~~~,~~~\psi^v_{K'\downarrow}(q)=\frac{1}{D^v_{+}}\begin{bmatrix}O\\O\\O\\t_0 q\\h^v_{+}\end{bmatrix}
\end{eqnarray}
%%%%%%
where $q=a_0(q_x+i q_y)$, $O$ is a two-component zero vector and other parameters read
%%%%%%
\begin{gather}
h^c_{\pm}=d_{\pm}-\sqrt{d^2_{\pm}+t_0^2 |a_0 {\bm q}|^2}~,
\nonumber\\
h^v_{\pm}=d_{\pm}+\sqrt{d^2_{\pm}+t_0^2 |a_0 {\bm q}|^2}~,
\nonumber\\
d_{\tau\times s}=\frac{\Delta+\lambda\tau s}{2}+b \beta |a_0 {\bm q}|^2~,
\nonumber\\
D^{c,v}_{\pm}=\sqrt{t_0^2 |a_0 {\bm q}|^2+h^{c,v}_{\pm}}~.
\end{gather}
%%%%%%
According to the general properties of the time reversal symmetry of a fermionic system ({\it i.e.} $\Theta^2=-1$), we can find the following relations
%%%%%%
\begin{gather}
\Theta \psi^c_{Ks}({\bm q})=i\psi^c_{K^\prime\bar s}(- {\bm q})~,\nonumber\\
\Theta \psi^c_{K^\prime\bar s}({\bm q})=-i\psi^c_{K s}(-{\bm q})~,\nonumber\\
\Theta \psi^v_{K s}({\bm q})=i\psi^v_{K^\prime\bar s}(-{\bm q})~,\nonumber\\
\Theta \psi^v_{K^\prime\bar s}({\bm q})=-i\psi^v_{K s}(-{\bm q})~.
\end{gather}
%%%%%%
Therefore, the matrix representation of the time reversal symmetry in this basis is given by
%%%%%%
\begin{equation}
\Theta\equiv i\begin{bmatrix}
O&O&O&M_{+}\\
O&O&M_{-}&O\\
O&-M_{-}&O&O\\
-M_{+}&O&O&O
\end{bmatrix}~,
\end{equation}
%%%%%%
in which
%%%%%%
\begin{equation}
M_{\pm}=\begin{bmatrix}
\frac{t_0^2|a_0 {\bm q}|^2-[h^c_{\pm}]^2}{[D^c_{\pm}]^2}
&
\frac{t_0^2 |a_0 {\bm q}|^2-h^c_{\pm}h^v_{\pm}}{D^c_{\pm}D^v_{\pm}}
\\
\\
\frac{t_0^2 |a_0 {\bm q}|^2-h^c_{\pm}h^v_{\pm}}{D^c_{\pm}D^v_{\pm}}
&
\frac{t_0^2 |a_0 {\bm q}|^2-[h^v_{\pm}]^2}{[D^v_{\pm}]^2}
\end{bmatrix}~,
\end{equation}
%%%%%%
It is easy to show that
%%%%%%
\begin{equation}
{\rm Pf}[\Theta]={\rm det}[M_{+}]{\rm det}[M_{-}]~,
\end{equation}
%%%%%%
By defining ${\rm det}[M_{\pm}]=[P_{\pm}( {\bm q})]^2$, we arrive at the following relation for the Pfaffian of the time reversal operator in the low-energy model.
%%%%%%
\begin{equation}
{\rm Pf}[\Theta]=\left[ P_{+}( {\bm q}) P_{-}( {\bm q})\right ]^2~.
\end{equation}
%%%%%%
Notice that to obtain the above relation for the Pfaffian following identities can be useful
%%%%%%
\begin{gather}
h^c_{\pm} h^v_{\pm} = -t_0^2|a_0 {\bm q}|^2~,
\nonumber\\
(D^c_{\pm} D^v_{\pm})^2 = 4 t_0^2 |a_0 {\bm q}|^2[d^2_{\pm}+t_0^2 |a_0 {\bm q}|^2]~.
\end{gather}
%%%%%%
%%%%%%
%\newpage
\section*{References}
\bibliographystyle{iopart-num}
\bibliography{Bibliography}
%%%%%%
\end{document}